\documentclass{article}

\usepackage{arxiv}

\usepackage[utf8]{inputenc} 
\usepackage[T1]{fontenc}    
\usepackage{hyperref}       
\usepackage{url}            
\usepackage{booktabs}       
\usepackage{amsfonts}       
\usepackage{amsmath}
\usepackage{nicefrac}       
\usepackage{microtype}      
\usepackage{lipsum}		
\usepackage{graphicx}
\usepackage{natbib}
\usepackage{doi}
\usepackage[affil-it]{authblk}

\usepackage[x11names]{xcolor}

\usepackage{hyperref}
\hypersetup{citecolor=DodgerBlue3, citebordercolor=DodgerBlue3, colorlinks=true}

\title{%
  Microbiomes Through the Looking Glass \\
  \large Linking
Species Interactions to Dysbiosis through a Disordered Lotka-Volterra Framework}

\author[1]{Jacopo Pasqualini}
\author[1]{Amos Maritan}
\author[2,3]{Andrea Rinaldo}
\author[4]{Sonia Facchin}
\author[4]{Edoardo Savarino}
\author[5,*]{Ada Altieri }
\author[1,6,*]{ \\ Samir Suweis}

\affil[1]{Dipartimento di Fisica “G. Galilei”  e \emph{INFN} sezione di Padova, Università di  Padova, Italy}
\affil[2]{Dipartimento di Ingegneria Civile, Edile e Ambientale \emph{ICEA}, University of Padova, \ Italy}
\affil[3]{\emph{EPFL}, Ecole Polytechnique Fedéralé Lausanne, 1015 Lausanne, Switzerland }
\affil[4]{Dipartimento di Scienze Chirurgiche, Oncologiche e  Gastroenterologiche \emph{DiSCOG}, University of Padova,  Italy }
\affil[5]{Laboratoire Matière et Systèmes Complexes (\emph{MSC}), Université Paris Cité, CNRS, 75013 \ Paris, \ France }
\affil[6]{Padova Neuroscience Center, University of Padova, Italy }
\affil[*]{Corresponding authors: \textsc{ada.altieri@u-paris.fr}, \textsc{samir.suweis@pd.infn.it} }

\renewcommand{\headeright}{ }
\renewcommand{\undertitle}{ }
\renewcommand{\shorttitle}{Microbiomes Through the Looking Glass}

\begin{document}

\maketitle

\begin{abstract}
The rapid advancement of environmental sequencing technologies, such as metagenomics, has significantly enhanced our ability to study microbial communities. The eubiotic composition of these communities is crucial for maintaining ecological functions and host health. Species diversity is only one facet of a healthy community’s organization; together with abundance distributions and interaction structures, it shapes reproducible macroecological states, i.e., joint statistical fingerprints that summarize whole-community behavior. 
Despite recent developments, a theoretical framework connecting empirical data with ecosystem modeling is still in its infancy, particularly in the context of disordered systems. Here, we present a novel framework that couples statistical-physics tools for disordered systems with metagenomic data, explicitly linking diversity, interactions, and stability to define and compare these macroecological states. By employing the generalized Lotka-Volterra model with random interactions, we reveal two different emergent patterns of species-interaction networks and species abundance distributions for healthy and diseased microbiomes. On the one hand, healthy microbiomes have similar community structures across individuals, characterized by strong species interactions and abundance diversity consistent with neutral stochastic fluctuations. On the other hand, diseased microbiomes show greater variability driven by deterministic factors, thus resulting in less ecologically stable and more divergent communities. Our findings suggest the potential of disordered system theory to characterize microbiomes and to capture the role of ecological interactions on stability and functioning.
\end{abstract}

\newpage

\section{Introduction}

Microbial communities are a fundamental reservoir of ecological functions and biological diversity. They are relevant for any environmental and host-associated ecosystem, ranging from soil to the human gut. In particular, the human gut microbiome, by interacting with the host's metabolism and immune system \cite{levy2016metabolites, barroso2015adaptive, barreto2023}, is a key regulator of human health. Dysbiosis is defined as an alteration in the composition of healthy microbiomes associated with the gastrointestinal tract \cite{lloyd2019multi,das2019homeostasis,nishida2018gut}.

Recently, stochastic non-interacting neutral \cite{zeng2015neutral,venkataraman2015application,sala2016stochastic,sieber2019neutrality}\cite{descheemaeker2020stochastic,grilli2020macroecological,zaoli2022stochastic}, and logistic models have been successfully used to describe empirical patterns such as species abundance distribution (SAD) and species presence/absence statistics in different types of communities. However, focusing on single-species properties \cite{azaele2016statistical,grilli2020macroecological} fails to characterize altered states of microbiome \cite{seppi2023emergent,pasqualini2024emergent}. In particular, recent studies indicate that gut dysbiosis is associated with shifts in microbial species interaction patterns. Network-level properties of species interactions -- such as the balance of positive and negative interactions, average interaction strength, and connectivity of the inferred interaction matrix -- have been shown to systematically differ between healthy and dysbiotic gut communities (see, for instance, \cite{bashan2016universality,seppi2023emergent}). Pairwise species interactions have been quantified in simplified in‑vitro consortia \cite{kehe2021positive,venturelli2018deciphering}; we assume that the same classes of interactions also operate -- albeit in a more complex form -- in the native gut microbiome.

For these reasons, several approaches have been proposed in recent years to infer microbial interactions \cite{faust2012microbial, xiao2017mapping,camacho2024sparse}. However, such inference protocols remain very challenging and problematic in many respects \cite{faisal2010inferring,angulo2017fundamental,tu2019reconciling,armitage2019sample,holt2020some}: a) the very high dimension of typical microbiome datasets and the lack of longitudinal (long-term) experiments; b) the time-dependent nature of microbial species interactions \cite{pacciani2021constrained}; c) abiotic factors such as resources, temperature, and pH, can introduce environmental filtering effects \cite{sireci2022environmental} and induce effective interactions;
d) experimental and technical challenges that introduce sampling effects, false positive species \cite{tovo2020taxonomic}, spurious correlation, and bias in the data \cite{weiss2016correlation,gloor2017microbiome,dohlman2019mapping}.
Therefore, even in a scenario where metagenomic samples are noise- and bias-free, reconstructing species-interaction networks of the underlying microbial dynamics remains a hard task.

An alternative, theoretically more elegant approach was pioneered by May \cite{may1972will}, proposing to use random matrices to model species interaction networks, i.e., each entry of the adjacency matrix is extracted at random from a given distribution. Given the impossibility of empirically measuring the interaction strengths, the advantage of such an approach is the reduction from $\sim S^2$ (if $S$ is the number of species) to just a few parameters (e.g., 2) used to parameterize the distribution \cite{allesina2012stability}.

\subsection{Insights from Disordered Systems Theory}

Although previous studies have largely focused on neutral theory and logistic models -- including variants like the Stochastic Logistic Model that account for environmental fluctuations -- they often neglect inter-species interactions, limiting their ability to reproduce macroecological patterns at a global scale. Moreover, higher-order correlations among species are known to generate non-trivial effects, such as the emergence of persistent fluctuations or chaotic dynamics due to the non-reciprocity of interactions.

In recent years, increasing attention has been devoted to studying population dynamics through the generalized Lotka-Volterra (gLV) equations, employing disordered systems techniques (also known as \emph{glassy}), such as replica and cavity methods and Dynamical Mean-Field Theory approaches, originally developed in the context of statistical physics \cite{bunin2017ecological,galla2018dynamically,biroli2018marginally,altieri2021properties, lorenzana2022well}. 
Indeed, due to the inherently high dimensionality of microbial datasets, random matrix theory and methods from disordered systems turn out to be particularly well suited. 
A striking feature of their application to ecological dynamics is that the resulting properties and dynamical regimes do not depend on species-specific details, provided that species are statistically equivalent under relabeling or time-averaging. Phase diagrams can thus be established in terms of a few effective parameters (e.g., the mean $\mu$ and the variance $\sigma^2$ of the random interactions strengths) \cite{barbier2018generic}. However, despite a range of very interesting results, this approach has remained confined mainly to purely theoretical domains (but see \cite{doi:10.1126/science.adg8488}) and has only recently been used to give an interpretation of controlled experiments with synthetic microbial communities \cite{hu2022emergent}.

\section{Results}

We now provide a \emph{proof of concept} of the applicability of a high-dimensional disordered setting to human microbiomes. 
Specifically in the following section, we introduce the disordered generalized Lokta-Volterra (dgLV) model, infer its parameters from healthy and unhealthy cohorts of gut microbiomes, and test whether the resulting interaction patterns and stability metrics discriminate healthy from diseased macro-ecological states. Finally, we will introduce quantitative metrics to define microbiome stability and estimate the contribution of distinct ecological forces to the dynamics.

\subsection{Disordered generalized Lotka-Volterra model}
The dgLV model describes the time evolution of the concentration abundances of a local pool of $S$ interacting species, i.e.,

\begin{equation}
    \frac{d N_i}{dt}= N_i \left[ \rho_i ( K_i - N_i) -\sum_{j, (j \neq i)} \alpha_{ij} N_j \right] + \eta_i(t) +\lambda \ ,
    \label{eq:LangevinGLV}
\end{equation}

where $N_i$ is the populations of species $i$-th, $K_i$ is its carrying capacity, and $r_i$ the growth rate where $\rho_i=\frac{r_i}{K_i}$ is a constant, which we will assume will not depend on the species, i.e. $\rho_i=\rho$. The coefficients $\alpha_{ij}$ are random i.i.d random variables with $\mathbb{E}[\alpha_{ij}]=\mu/S$ and $\mathbb{E}[\alpha_{ij}]^2 -\mathbb{E}^2[\alpha_{ij}]=\sigma^2/S$. We also include an immigration species-independent rate $\lambda$, which will be treated as a reflecting wall mathematically regularizing the problem, and a demographic noise term \cite{altieri2021properties} $\eta_i$, with $\langle \eta_i(t)\eta_j(t')\rangle=2 N_i T \delta_{ij} \delta(t-t')$, and where $\delta$ denotes the Dirac delta and $T$ sets the scale of noise intensity. For notational purposes, we also introduce its inverse $\beta=T^{-1}$. Finally, we require that the interactions are symmetric, i.e., $\alpha_{ij}=\alpha_{ji}$. We can thus map this problem into an equilibrium thermodynamic one that is exactly solvable.  
As shown in the Appendix, section S1, it is possible to justify this symmetric assumption and a linear dependence between the growth rate and the carrying capacity, namely $r_i=\rho K_i$, by considering the quasi-stationary approximation of the MacArthur consumer-resource model. In the following, we set $\rho=1$ (following also \cite{biroli2018marginally}).

The presence of a noise term in Eq. (\ref{eq:LangevinGLV}) allows us to write the Fokker-Planck equation of the system and study its stationary solution (see \cite{altieri2021properties, Altieri2022} for a detailed derivation in a similar Hamiltonian formalism). Here, we adopt the It$\hat{\text{o}}$ prescription for the stochastic dynamics in such a way as to prevent species resurgence by noise.

The replica formalism, a well-known technique in disordered systems, comes into play and allows us to derive a non-interacting Hamiltonian corresponding to the dgLV symmetric interactions $\alpha_{ij}$ (see Methods and Appendix, Section S2, for a complete derivation). In simple words, instead of trying to solve the problem for one random setup, the formalism considers many replicas of the system and averages their behaviors. This approach helps to smooth out the randomness and reveals the typical behavior of the system. We also assume a single equilibrium scenario, known as \emph{replica-symmetric} (RS) regime. Although the SADs of empirical microbial communities display fat tails, a feature more compatible with the multi-attractors phase of the asymmetric 1RSB \cite{mallmin2024chaotic} case or with the gLV with time-dependent disordered interactions \cite{suweis2024generalized} (\emph{annealed} version), we consider this simplification as it is the regime in which we can obtain explicit analytical relations between the model parameters and the data and where model inversion is feasible. 
By employing a cavity argument \cite{mezard-montanari, altieri2024introduction}, one can indeed analytically derive the SAD of the model (see Methods and Appendix, Section S4):

\begin{equation}\label{eq:cavity_sad}
    p(N|\zeta) \propto N^{\nu-1} \exp \Biggr \lbrace - \beta \left ( \frac{m}{2}N^2 - \zeta N \right ) \Biggr \rbrace ,
\end{equation}

where the auxiliary variable $\zeta = K-\mu h+\sqrt{q_0}\sigma z$ takes the disorder interactions into account; $z$ is a standardized Gaussian variable, $\nu=\beta \lambda > 0$. We introduce, respectively, the mean abundance $h=\overline{ \langle  N \rangle}$, the self-overlap $q_d=\overline{\langle  N^2 \rangle}$, and the overlap $q_0=\overline{ \langle N \rangle^2}$, which we will collectively refer to as \emph{order  parameters} in the following sections. Specifically, $q_0$ measures the similarity between two different configurations at stationarity of the system with two different disorder realizations, while $q_d$ measures the similarity of two stationary configurations generated with the same disorder realization. Finally, we introduce a constant that, inspired by the Field Theory jargon, we dub as mass  $m=1-\beta\sigma^2(q_d-q_0)>0$ for the theory to be consistent \cite{biroli2018marginally}.

The RS ansatz can also be characterized by its stability to external perturbations, such as the external temperature, the immigration rate, or the interaction heterogeneity. To investigate the stability of the RS phase, we consider the Hessian matrix of the theory by performing a harmonic expansion of the replicated free energy, as originally pointed out in \cite{de1978stability}. When the leading eigenvalue of the Hessian matrix, the so-called \emph{replicon mode}, goes below zero, unstable equilibria appear: either the system moves towards a One-step Replica Symmetric Breaking (1-RSB) phase with multiple locally stable minima, or it develops a marginal full-RSB phase, leading to a hierarchical structure of states and therefore to an extremely rough landscape (also known as \emph{Gardner amorphous-like phase}). In other words, the RS ansatz no longer represents a thermodynamically stable phase in the latter cases.

Based on standard calculations in disordered systems, the  \emph{replicon} mode $\mathcal{R}$ can be analytically evaluated and reads\footnote{The replicon eigenvalue refers to a particular type of fluctuation around the saddle-point (mean-field) solution in the replica framework. When diagonalizing the Hessian matrix of the replicated free energy, fluctuations split into three sectors: longitudinal, anomalous, and replicon. The replicon mode is the most sensitive to criticality signaling – by its vanishing trend – the emergence of many nearly-degenerate states. It essentially describes how ‘soft’ the system is to microscopic rearrangements in configuration space.}:
\begin{equation}
    \mathcal{R} =  \left ( \beta \sigma \right )^2  \left ( 1 - \sigma^2 \overline{ \langle \left( \frac{\partial N}{\partial \xi }  \right)^2 \rangle}\right ) = \left ( \beta\sigma \right )^2  \left ( 1- \left ( \beta \sigma \right )^2 \overline{ \left(\langle N^2 \rangle - \langle N \rangle^2 \right)^2} \right), \
    \label{replicon}
\end{equation} \\
where $\frac{\partial N}{\partial \xi}$ is the species response to an external perturbation, keeping track of non-extinct species only. For more details, see S3 in the Appendix.

\subsection{Data Through the Glass} 
We now consider cross-sectionally sampled gut microbiomes of two different cohorts: one of healthy and another of diseased individuals. Here we focus on chronic inflammation syndromes, but, in principle, what we present holds for groups of phenotypically distinct populations. We can idealize the samples of each group as generated from the stationary distribution of the dgLV Eq. (\ref{eq:LangevinGLV}), with \emph{different} realizations of the disorder $\boldsymbol{\alpha}$, but with shared $\mu$ and $\sigma$ (see Material and Methods). The approach we propose considers all samples in the same group as coming from the same statistical ensemble. On the other hand, it distinguishes ecological regimes characterized by distinct phenotypes through different statistics (e.g., dissimilar $\mu$ and $\sigma$) of the random species interactions $\boldsymbol{\alpha}$. We aim to test the hypothesis that each cohort will be described by a different set of ecological parameters, precisely with $\boldsymbol{\theta}=(\mu,\sigma,T,\lambda)$. 
\vspace{0.3cm}

\begin{figure}[h]
    \centering   
    \includegraphics[width=0.99\linewidth]{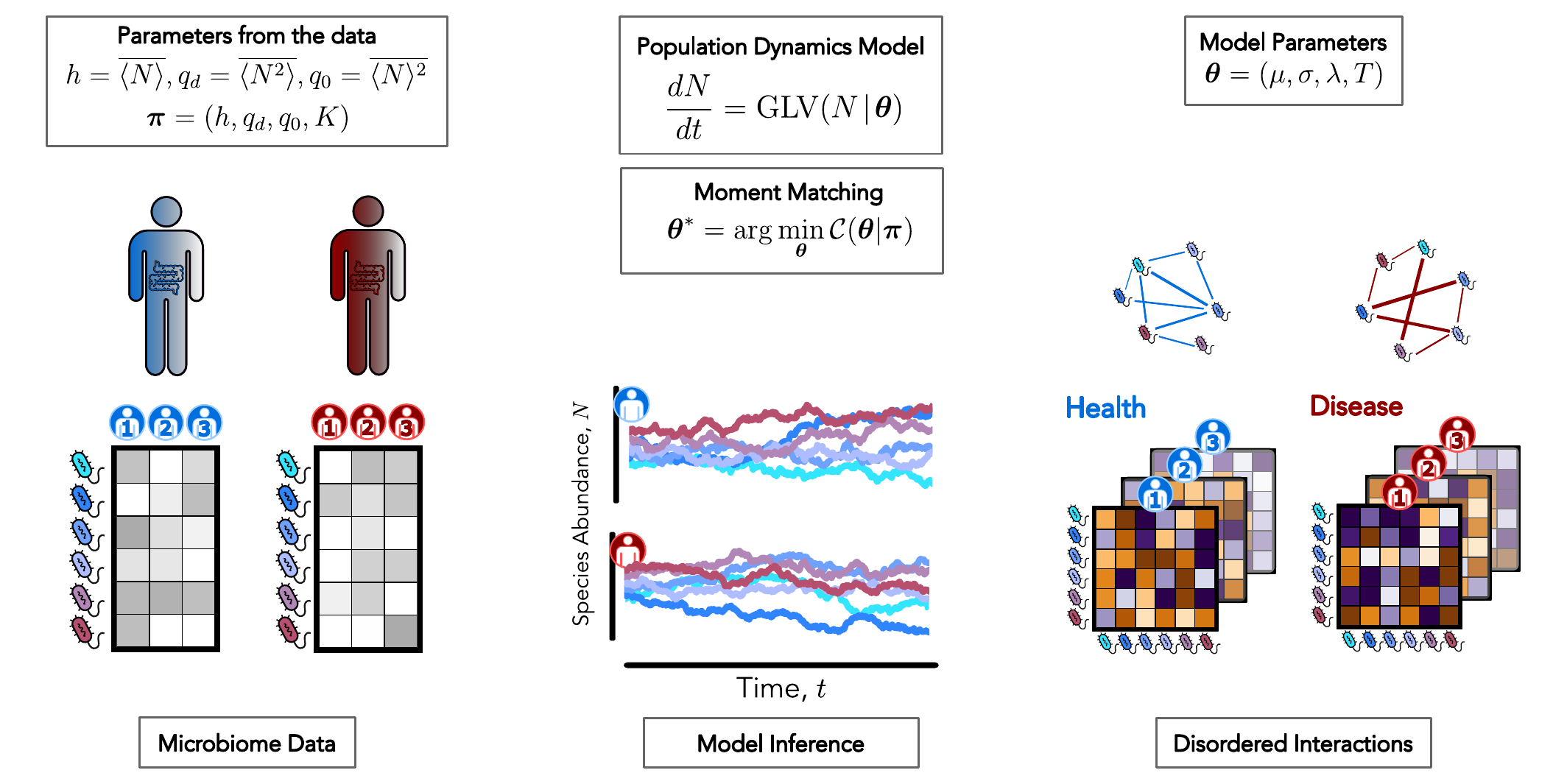}
    \caption{The inference protocol of the dgLV generative model is performed by a moment matching optimization procedure. We aim to infer the free parameters $\boldsymbol{\theta}=(\mu,\sigma,T,\lambda)$ -- shown on the right --  so that to match the mean abundance, higher-order correlations between species abundances, and the average carrying capacity for the two cohorts, i.e. $(h, q_0, q_d, K)$ -- on the left. This procedure allows us to extract ecological dynamics information for cross-sectional data of healthy (blue) and diseased (red) microbiomes, which are conceptualized as independent disorder realizations. }
   \label{fig:one}
\end{figure}

To calculate the order parameters $(h, q_d, q_0, K)$ from the data, we thus need to specify how we perform the ensemble and the disorder averages empirically (see Figure \ref{fig:one}). 
Since time series data are rarely available, we rely on an effective mean-field description and estimate the ensemble average by averaging over species (that is, different species are realizations of the same underlying stochastic process \cite{azaele2016statistical}, known as neutral hypothesis), i.e.,
$ \langle \ \cdot \  \rangle \sim \frac{1}{S}\sum_{\textit{species}}\cdot$. Then, we assume that the average over the disorder can be computed as a sample average. In other words, within a given phenotype (healthy/unhealthy), each measured microbiome configuration is a sample from the stationary distribution of the dgLV model, with a given realization of the disorder, i.e., $ \overline{ \ \cdot \ } \sim \frac{1}{R}\sum_{\textit{samples}}\cdot$, where $R$ is the number of samples and typically $R\gg1$. Moreover, due to our limited knowledge of the fine details governing the interactions in each microbiome, it is reasonable to assume that all communities with a given macrostate -- a point in the ($h$, $q_0$, $q_d$, $K$) space -- experience the same demographic noise $T$, immigration rate $\lambda$, and disorder parameters $\mu$ and $\sigma$. Eventually, we can compute the order parameters from the data as:

\begin{equation}
\begin{split}
    h & = \overline{\langle N \rangle } = \frac{1}{R}\sum_{a=1}^{R} \left ( \frac{1}{S_a}\sum_{j=1}^{S} N_{j,a} \right ) \ , \\
    q_d & =  \overline{\langle N^2 \rangle } = \frac{1}{R}\sum_{a=1}^{R} \left ( \frac{1}{S_a}\sum_{j=1}^{S} N^2_{j,a} \right ) \ , \\
    q_0 & =  \overline{\langle N \rangle^2 } = \frac{1}{R}\sum_{a=1}^{R} \left ( \frac{1}{S_a}\sum_{j=1}^{S} N_{j,a} \right )^2 \,
\end{split}\label{eq:op2data}
\end{equation}
where $N_{j,a}$ represents the population density of species $j$ in sample $a$.

To evaluate the carrying capacities $K_i$ from the data, as done in most of the works using dgLV equations \cite{bunin2017ecological,biroli2018marginally,altieri2021properties,suweis2024generalized,mallmin2024chaotic}, we assume that in each cohort all species are characterized by the same carrying capacity
$K_i=K$. Here, we define $K$ as the average of the maximum relative abundances of species across the available samples for each cohort, so that $K=\frac{1}{S}\sum_{j=1}^{S} \max_{a} N_{j,a}$.

\vspace{0.5cm}

\begin{figure}[h]
    \centering   
    \includegraphics[width=0.9\linewidth]{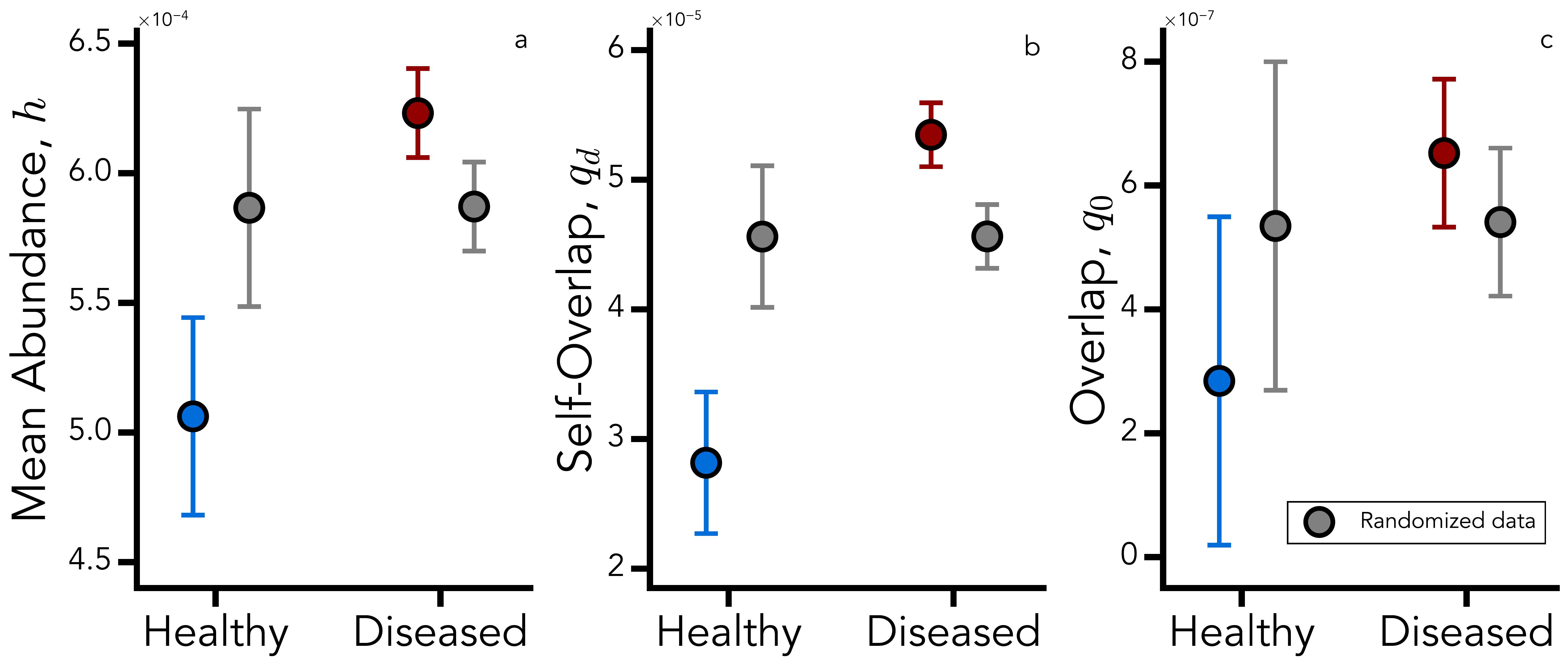}
    \caption{Order parameters inferred from the data using Eqs.~(\ref{eq:op2data}). Panel a) shows $h$, b) $q_d$, and panel c) $q_0$ in healthy (blue) and unhealthy (red) cohorts. The gray points show the values for the corresponding null models where sample labels have been randomized. Standard deviations have been computed over 5000 realizations.}
   \label{fig:two}
\end{figure}

Note also that, from metagenomics, we only have access to compositional data for species abundances \cite{pasqualini2024emergent}. This is crucial to properly treat different samples and end up with a consistent analysis. 
Moreover, this approach allows us to give an ecological interpretation of some of the order parameters. In particular, because of the compositionality of the data, we have that $h=\overline{S^{-1}}$ and $q_0=\overline{S^{-2}}$, while $q_d$ provides information on pairwise products between species abundances within each replica. Figure \ref{fig:two} shows the values of such order parameters between healthy and unhealthy cohorts for two randomized instances. In the data, we observe a systematic difference between healthy and unhealthy cohorts, pointing to a higher average local diversity in healthy samples (panels a, c), as also observed in \cite{pasqualini2024emergent}. Panel b highlights a higher $q_d$ in unhealthy patients, a signature of the weakening species interactions in those samples. We will further investigate this aspect by inferring the species interactions in both cohorts.

From the modeling point of view, it can be rigorously shown that constraining the abundances to be normalized to one corresponds to introducing a Lagrange multiplier in the expression of the free energy. However, this only acts on the linear term, contributing to shifting the mean of the random interactions, $\mu$, but does not affect the heterogeneity $\sigma$, and therefore the phase diagram overall. For more details, we refer the interested reader to the Appendix (end of Section S2) along with \cite{altieri2021properties}. In \cite{altieri2021properties} the authors also showed how a global normalizing constraint on species abundances reflects in a one-to-one mapping with the random replicant model \cite{Biscari1995}\footnote{The replicator equations originally introduced by \cite{Diederich1989} and recast within the replica formalism \cite{Biscari1995, altieri2021properties} describe the dynamics of an ensemble of replicants that evolve via random couplings.}.

\subsection{Species interactions patterns characterize the state of microbiomes}%

We thus collect all the parameters estimated from the data in a vector $\boldsymbol{\pi}=(h,q_d,q_0,K)$.
As we will better detail in the Methods, we develop a moment matching inference algorithm to infer the model parameters $\boldsymbol{\theta}$, as depicted in Figure \ref{fig:one}. The idea of the method is to introduce a cost function $\mathcal{C}(\boldsymbol{\theta}|\boldsymbol{\pi})$, representing a total relative error for some self-consistent equations. If the parameters $\boldsymbol{\theta}$ are such that the right part of the self-consistent equation equals the left part, the problem is considered solved. Because the landscape associated with this cost function presents several minima, we perform multiple optimization procedures to collect an ensemble of possible solutions, from which we retain the top 30.
First, we find that different solutions $\boldsymbol{\theta}^*$ of the optimization problem provide ecological insights into the underlying microbiome populations. 

\begin{figure}[h]
\centering
\includegraphics[width=0.95\linewidth]{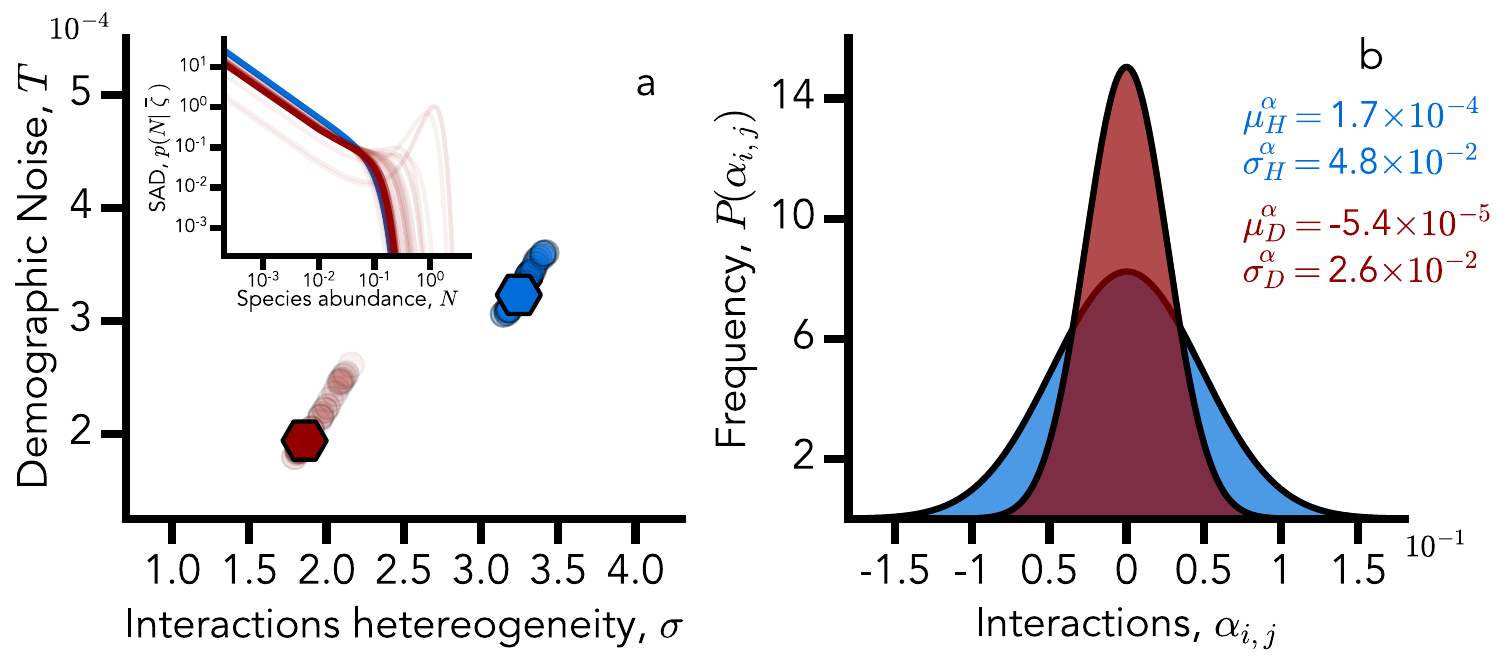}
   \caption{Panel a: inferred $T$ (demographic noise strength) and $\sigma$ (interactions heterogeneity) for healthy (blue) and diseased (red) microbiomes are clustered. Darker dots correspond to better solutions (i.e., solutions with a lower value of the cost function $\mathcal{C}$), while the two points with hexagonal markers correspond to the best two (healthy and diseased, respectively) solutions. In the first panel inset, we also show (in log-log scale) the SADs corresponding to each solution. To have a more concise representation, we present each SAD fixing the disorder to its average $\overline{\zeta} = K - \mu h$. Panel b: the probability density function of the inferred interactions $\alpha_{i,j}$ for healthy (blue) and diseased (red) microbiomes. Dysbiosis reduces the heterogeneity of the interaction strengths. The quantities reported in the legend are the average and standard deviation of $\alpha_{i,j}$. They have are calculated as $\mu^{\alpha}_{X} = \mu_{X} / S_{X}$ and $\sigma^{\alpha}_{X}= \sigma_{X} / \sqrt{S_{X}}$, where $S_{X}$ is the species pool size, estimates as the set of all the observed species in a dataset, $X$. $X$ can be healthy ($H$) or diseased ($D$).}
   \label{fig3}
\end{figure}

As originally predicted in \cite{altieri2021properties}, among all the parameters that define Eq. (\ref{eq:LangevinGLV}), the only ones relevant for reproducing the theoretical phase diagram are the amplitude of demographic noise and the heterogeneity of interactions. The mean interaction strength, provided it is sufficiently positive, does not play a significant role.
This prediction is fully confirmed by the inference procedure applied to the two microbiome datasets, allowing us to identify a universal signature that distinguishes healthy from unhealthy states. Figure \ref{fig3}a shows, indeed, that inferred noise ($T$) and interaction heterogeneity strength ($\sigma$) for healthy and diseased microbiomes are clustered in the two-dimensional plane. 

In particular, the SAD for the healthy cohort is robust among the different solutions of the inference procedure, as depicted by the superposition of the different curves in the inset of Figure \ref{fig3}A. On the other hand, SADs inferred from unhealthy patients have high sensitivity to different solutions. In particular, some of them display a mode for high-abundance species (light red lines in Figure \ref{fig3}a), a signature of dominant strain in the gut. 
Consistently, the distribution of the interactions $P(\alpha_{i,j})$ generated through the inferred parameters $\mu$ and $\sigma$ is different between healthy and diseased cohorts, giving a distinct pattern of interactions (see Figure \ref{fig3}b), a result that is compatible with that found by Bashan and collaborators \cite{bashan2016universality}. Remarkably, we find that dysbiosis reduces the heterogeneity of interaction strengths, a result also observed when taking correlations as a proxy for interactions \cite{seppi2023emergent}. 

\begin{figure}[h]
\includegraphics[width=1.1\linewidth]{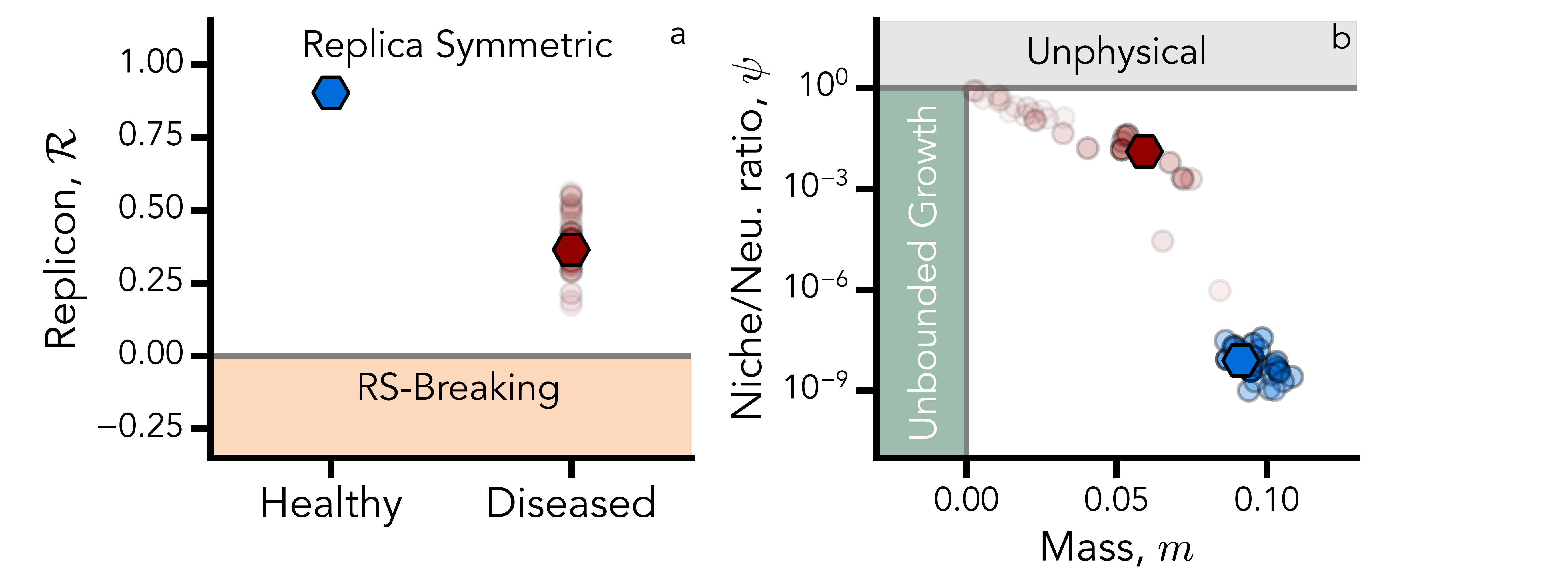}
   \caption{Panel a: The replicon eigenvalue corresponding to each solution of our optimization procedure (shaded dots). The solid hexagon represents the replicon corresponding to the best solutions that minimize the error in predicting the order parameters of the theory (minimum $\mathcal{C}$). The two investigated microbiome phenotypes (healthy in blue, diseased in red) are significantly different. In particular, diseased microbiomes are closer to the marginal stability of replica-symmetric ansatz (grey horizontal line). Panel b: Solutions of the moment-matching objective function are shown as a function of $\psi$ and $m$, which in turn depend on the SAD parameters (see main text). Healthy (blue) and diseased (red) microbiomes appear to be clustered. Therefore, distinct ecological organization scenarios (strong neutrality/emergent neutrality) emerge. Darker dots correspond to solutions with lower values of the cost function, while hexagonal markers correspond to the two best solutions.}
   \label{fig4}
\end{figure}

We then assess how close the inferred $\sigma$ and $\beta=1/T$ (a.k.a. inverse temperature in a statistical physics approach) are to the critical replica symmetry-breaking (RSB) line of the dgLV ($\mathcal{R}=0$), evaluated by keeping all the other parameters constant (see Methods). 
We find again that the replicon values $\mathcal{R}$ corresponding to each solution of our optimization protocol are significantly different for the two investigated microbiome phenotypes (see Figure \ref{fig4}a). In particular, diseased microbiomes are closer to marginal stability within the replica-symmetric ansatz \cite{altieri2021properties, mezard1987spin, de1978stability}. 
Furthermore, by investigating the shape of the SAD given by Eq. (\ref{eq:cavity_sad}), we can estimate the ratio between niche (represented by species interaction) and neutral (represented by birth/death and immigration) ecological forces, which can be captured by the quantity $\psi$ \cite{wu2021understanding}. It detects the emergence of peaks in the species abundance distribution as a hallmark of niche processes (see Appendix 2). 

Inspired by field-theory arguments (see Methods and Appendix, Section S2), we can call \emph{mass} of the theory the $m$ parameter, as defined above in Eq.~(\ref{eq:cavity_sad}). In classical and quantum field theory, the particle-particle interaction embedded in the quadratic term is typically referred to as a mass source. In our context, $m=1-\beta\sigma^2(q_d-q_0)$ captures quadratic fluctuations of species abundances, as also appearing in the expression of the leading eigenvalue of the stability matrix.
When $m\rightarrow0$, the analytical order parameters diverge and the system enters the unphysical regime of unbounded growth. As such, the \emph{mass} term can be considered a complementary stability measure, capable of capturing the transition to the unbounded growth regime.

In the model, two kinds of effects compete to shape the community structure. On the one hand, we have niche effects, encoded in disordered interactions and thus tracked by the parameters $\mu$, $\sigma$, and $K$. Their overall effect is selective and tends to concentrate the SAD around the typical abundance value. On the other hand, we have neutral effects encoded in the stochastic dynamics and immigration, governing the low abundance regime of the SAD. When the demographic noise amplitude is stronger than immigration ($\nu<1$, as in our case), the SAD exhibits a low-abundance integrable divergence. In the opposite scenario, for $\nu>1$, there is no divergence and the SAD is modal. Since interactions are random, the probability of observing an internal mode can be estimated as the fraction of SADs realizations having non-trivial solutions to the stationary point equation. Such a quantity, dubbed as the niche-neutral ratio, can be analytically evaluated:

\begin{equation}
    \psi = \frac{1}{2}  \text{Erfc} \left ( \frac{ \zeta^*+\overline{\zeta }}{\sqrt{2} \sigma_{\zeta}} \right ) + \frac{1}{2} \text{Erfc} \left ( \frac{\zeta^*-\overline{\zeta }}{\sqrt{2} \sigma_{\zeta}} \right )  \ ,
\end{equation}

where $\zeta^*=\sqrt{\frac{4(1-\nu)m}{\beta}}$ and $\overline{\zeta}=K-\mu h$. When $\psi \approx 1$, niche and neutral forces give comparable contributions to the dynamics, as both low abundance divergence and a finite abundance mode coexist in the SAD. Finally, if the typical abundance diverges, we enter the unbounded growth phase, which means that the mass $m$ and the niche-neutral ratio $\psi$ are not independent, as suggested by the analytical expression for $\psi$. For an exhaustive derivation of this result, see Appendix 2. With the obtained model parameters, we are able to evaluate $m$ and $\psi$ for healthy and diseased microbiomes. Also in this case healthy and diseased microbiomes are visibly clustered, as shown in Figure \ref{fig4}. Unhealthy microbiomes turn out to be closer to the unbounded growth phase, and the niche-neutral ratio is larger by five orders of magnitude than the healthy case $\psi_D \approx 10^{5} \psi_H $. This leads us to argue that selective pressure is way larger in diseased states, while in the healthy one, birth and death effects are the key drivers of the dynamics. These results are also confirmed by the SAD shapes in the inset of Figure \ref{fig3} (panel a).

In summary, in the Result section we demonstrate that: (i) the inference pipeline robustly recovers demographic noise and interaction heterogeneity by calculating $h$, $  q_0$ and $q_d$ from the data; (ii) These parameters cluster by health status and (iii) diseased microbiomes lie closer to the replica-symmetry-breaking threshold, indicating reduced ecological resilience.

\section{Discussion and Conclusions}
In our exploration of the gut microbiome through the lens of disordered systems and random matrix theory, we have proposed a connection between the theoretical framework of disordered systems and practical analyses of environmental microbiome data. In particular, we have characterized healthy and unhealthy gut microbiomes using the disordered generalized Lotka-Volterra model (dgLV) for population dynamics.
We now interpret our theoretical findings in a biological context, contrast them with previous work, and outline limitations and future directions.

The first major result of our work suggests a different role for the various ecological forces shaping the human gut microbial community. In this sense, the niche-neutral ratio $\psi$, highlights the different roles of interactions in healthy and diseased microbiomes. In the healthy case, neutral forces, such as random birth and death of individuals, characterize the dynamics, making configurations corresponding to this state alike. On the contrary, in diseased microbiomes, disordered, sample-specific interactions are the dominant ecological force, making individual realizations differ significantly from one another. An ecological interpretation of our findings suggests that healthy microbiomes are governed primarily by demographic stochasticity, reflecting a quasi-neutral regime characterized by similar community structures across individuals. Conversely, microbiomes from diseased patients exhibit significantly greater variability, suggesting that deterministic ecological factors -- such as weakened species interactions -- override neutrality, leading to structural instability and distinct microbial compositions. This observation aligns with the "Anna Karenina principle" \cite{pasqualini2024emergent,ma2020testing} holding for gut microbiomes, which can be phrased as: "All healthy gut microbiomes are alike; each unhealthy gut microbiome is unhealthy in its own way”. Supporting this interpretation, our analysis of the replicon eigenvalue $\mathcal{R}$ shows that the healthy state is associated with pronounced stability to external perturbations. The unhealthy state, instead, being closer to the RSB line, exhibits diminished stability and, consequently, reduced robustness against external perturbations. 

Our study also sheds light on the role of demographic noise within the context of the dgLV model. In the limit $T\rightarrow0$, the SAD transitions to a truncated Gaussian, with a prominent peak at zero reflecting the fraction of extinct species. In contrast to a zero-noise scenario, where species extinctions are observed, the inclusion of demographic noise in the dgLV model suggests a picture where no species goes extinct, supporting the \emph{everything is everywhere} hypothesis \cite{grilli2020macroecological,pigani2024}. In other words, within this framework, zeros in the data are due to sampling effects and not due to local species extinctions \cite{grilli2020macroecological,pasqualini2024emergent}.

A notable limitation of our study lies in the discrepancy between the empirical Species Abundance Distribution (SAD) observed in the data and the theoretical distribution predicted by the quenched dgLV model \cite{suweis2024generalized}. While empirical data showcase a diverse range of species abundance, following a power-law distribution, the model predictions tend to exhibit exponential decay in SAD tails. This mismatch underscores the need for further refinement of the model to accurately capture the nuanced patterns observed in real-world data. For example, it has recently been proposed that introducing an annealed disorder\footnote{Unlike the \emph{quenched} approximation, the \emph{annealed} version assumes that random couplings are not fixed but rather fluctuate over time, with their covariance governed by independent Ornstein–Uhlenbeck processes.} can generate SADs that more closely resemble the empirical ones \cite{suweis2024generalized}. Another possibility is to set the dgLV model parameters in such a way as to reproduce the multi-attractor phase. In fact, in this region, the SADs display a more heterogeneous shape \cite{pirey-PRX}. We plan to explore this follow-up direction by combining one-step replica-symmetry-breaking (1RSB) computations in the multiple-attractor phase with Dynamical Mean-Field Theory analysis for the asymmetric interaction case.
This latter approach is particularly well suited for studying inherently non-equilibrium dynamics and for extending the framework to systems subject to environmental fluctuations in addition to demographic noise.

Another related limitation is the challenge of generating species abundance samples from the dgLV model that mirrors the statistical properties of the observed empirical data. In our current framework, each microbiome sample could be extracted from $p(N|\zeta)$, where $\zeta$ is a realization of the disorder. However, $p(N|\zeta)$ near $N\rightarrow0$ presents a power-law exponent $\nu-1$, with $\nu_{H,D}\approx 10^{-3}$. This results in numerical instabilities and dominates the sampling process, posing difficulties in generating representative synthetic samples. Moreover, while microbiome data are inherently compositional \cite{pasqualini2024emergent}, the dgLV model species populations $N_i$ are positive real numbers. However, as already noted, it can be shown that introducing a normalization $\tilde{N}_i\rightarrow N_i/\sum_j N_j$ in such equations does not change the structure of the proposed solutions \cite{altieri2021properties} and therefore should not affect the conclusions of our work.


In conclusion, our work proposes a bridge between theory and data, particularly in refining the theoretical models to better align with empirical observations and in exploring the nuances of species abundance distributions within the microbiome context. 
Moreover, the integration of other forms of environmental variability and species-specific traits could provide a more holistic view of ecological dynamics, as also proposed by the \emph{One Health-One Microbiome} framework \cite{tomasulo}.

Overall, our study builds a quantitative link between metagenomic data and the disordered gLV framework, revealing how dysbiosis alters gut species interaction networks. By doing so, it lays the groundwork for more advanced, mechanistically informed models to better interpret and ultimately manage complex microbial ecosystems.
\vspace{0.5cm}

\section{Acknowledgments}
We thank Silvia De Monte for insightful discussions.
S.S. acknowledges Iniziativa PNC0000002-DARE - Digital Lifelong Prevention. A.M. acknowledges financial support under the National Recovery and Resilience Plan (NRRP), Mission 4, Component 2, Investment 1.1, Call for tender No. 104  by the Italian Ministry of University and Research (MUR), funded by the European Union – NextGenerationEU – Project Title “Emergent Dynamical Patterns of Disordered Systems with Applications to Natural Communities” – CUP 2022WPHMXK - Grant Assignment Decree No. 2022WPHMXK adopted on 19/09/2023 by the Italian Ministry of Ministry of University and Research (MUR).
A.A. acknowledges the support received from the Agence Nationale de la Recherche (ANR), under the grant  ANR-23-CE30-0012-01 (project “SIDECAR”). 

\section{Author contributions}
S.S. and A.A. designed and supervised research, J.P. performed research, E.V.S. and S.F. provided data filtering criteria; all the authors wrote the paper.



\newpage

\section*{Material and Methods}
\vspace{0.3cm}

\subsection*{Microbiome dataset and code}

We have selected gut microbiome data from three studies \cite{franzosa2019gut,lloyd2019multi,mars2020longitudinal}, focusing on inflammatory syndromes of the gastrointestinal tract (Crohn's Disease, Ulcerative Colitis, and Irritable Bowel Syndromes). 
Considering all the available metadata, we have selected the patients less affected by possible perturbing factors, such as drugs. Finally, our dataset consists of $R_{\text{Healthy}}=91$ shotgun metagenomic samples from healthy control individuals and $R_{\text{Diseased}}=202$ shotgun metagenomic samples. All metagenomic preprocessing and reads classification are extensively described in \cite{pasqualini2024emergent}. Species abundance profiles from metagenomic data and the parameters values obtained from the moment matching inference  are available at the associated \href{https://zenodo.org/records/11934376}{Zenodo page}. 

All the scripts implementing the moment matching inference and the jupyter notebooks to generate the figures are available at the associated \href{https://github.com/jacopopasqualini/MicrobiomesGlass}{GitHub page}.

\subsection*{Free-energy landscape exploration: replica formalism}

In the case of symmetric interactions -- corresponding to conservative forces in the dynamics -- a one-to-one mapping between the multi-species dynamics and a thermodynamic formalism can be safely worked out. The first step consists of writing the Fokker-Planck equation in the presence of a white Gaussian noise defined by a zero mean and a variance of amplitude $2T$. All technical details, leveraging a Fokker-Planck derivation, can be found in \cite{Altieri2022} for a similar pairwise interacting model, but a different self-regulation term accounting for non-logistic behavior (\ref{eq:LangevinGLV}).

Once the (quenched) disordered Hamiltonian of the model is obtained, we can resort to techniques known in statistical physics of disordered systems, such as the replica and cavity methods \cite{mezard-montanari, mezard1987spin, zamponi2010}.
The replica trick, in particular, allows us to handle disordered quantities, such as the free energy and the partition function, which would be otherwise unaffordable (see Appendix below). 

We summarize the main findings here along with the expressions of the (RS) order parameters of the model, $(h,q_d,q_0)$. The three expressions below, originally obtained in \cite{altieri2021properties}, have offered the starting point of this work, allowing for a thorough comparison with the same order parameters measured from metagenomic data. 
Their analytical expressions are self-consistently determined by the system of equations:
\begin{equation}
\begin{split}
& h= \int \mathcal{D} z \left ( \frac{ \int_{0}^{\infty} e^{-\beta \mathcal{H}_{\text{RS}}(N; q_d, q_0 h,z)} \; N}{ \int_{0}^{\infty} d N e^{-\beta \mathcal{H}_{\text{RS}}(N; q_d, q_0, h,z)}}\right )=\overline{ \langle  N \rangle} , \\
& q_d= \int \mathcal{D} z \left(\frac{ \int_{0}^{\infty} d N e^{-\beta \mathcal{H}_{\text{RS}}(N; q_d, q_0, h,z)} \; N^2}{ \int_{0}^{\infty} d N e^{-\beta \mathcal{H}_{\text{RS}}(N; q_d, q_0, h,z)}} \right) = \overline{\langle  N^2 \rangle} \ , \\ 
& q_0= \int \mathcal{D} z \left(\frac{ \int_{0}^{\infty} d N e^{-\beta \mathcal{H}_{\text{RS}}(N; q_d, q_0, h,z)} \; N}{ \int_{0}^{\infty} d N e^{-\beta \mathcal{H}_{\text{RS}}(N; q_d, q_0, h,z)}} \right)^2  = \overline{ \langle N \rangle^2}  
\label{eq:sp}
\end{split}
\end{equation}
where the calligraphic notation stands for the Gaussian integration $\mathcal{D}z \equiv \int \frac{dz}{\sqrt{2 \pi}} e^{-z^2/2}$. In other words, the external average is equivalent to averaging over the quenched disorder $\overline{\cdot}$, whereas the most internal one -- over the continuous variable $N$ -- is interpreted as a thermal average over the single-equilibrium Hamiltonian $\mathcal{H}_\text{RS}(N;q_d,q_0,h)$. The latter is denoted by $\langle \cdot \rangle$.
See Appendix (Section S2) for more details.

As long as the system of Eqs. (\ref{eq:sp}) admits physically reasonable solutions, we might claim that the RS ansatz safely holds. This condition is nevertheless necessary but not sufficient because the stability of the RS solution must also be checked. It therefore requires studying the Hessian matrix of free energy and diagonalizing it on a suitable subspace, called \emph{replicon}, $\mathcal{R}$.
The main outcome of this computation is captured by Eq. (\ref{replicon}) of the main text. 
The averaged difference describes the fluctuations between the first and second moments of the species abundances within one state, namely between the diagonal value $q_d$ and the off-diagonal contribution $q_0$ of the overlap matrix. 
Detecting a vanishing value of $\mathcal{R}$ corresponds to the appearance of a marginally stable RS solution (see Section S3 of the Appendix).

\subsection*{Moment matching inference}
The parameters $h$, $q_0$, and $q_d$ can be self-consistently determined through the saddle point of the dgLV free energy in Eqs. (\ref{eq:sp}) (see also \cite{altieri2021properties}). We thus aim to estimate which set of model parameters (i.e. $\boldsymbol{\theta}=(\mu,\sigma,T,\lambda)$) will generate values of the order parameters ($h$, $q_d$, $q_0$) matching those directly estimated from the data. The solution of such an inference problem may not be unique or exact. We have thus developed an optimization algorithm so to find a pool of possible solutions that minimize the difference between the order parameters estimated by the model and those directly obtained from the data.
To infer the parameters $\boldsymbol{\theta}$, we can thus use the self-consistent equations for the order parameters and solve the inverse problem to find the dgLV parameters that match the empirical observations. For each self-consistent equation, we can introduce a  relative error $\delta H(\boldsymbol{\theta}|\boldsymbol{\pi}) = (H(\boldsymbol{\theta}|\boldsymbol{\pi})-h)/h$, 
$    \delta Q_d(\boldsymbol{\theta}|\boldsymbol{\pi}) = (Q_d(\boldsymbol{\theta}|\boldsymbol{\pi})-q_d)/q_d $ and 
$    \delta Q_0(\boldsymbol{\theta}|\boldsymbol{\pi})  =  (Q_0(\boldsymbol{\theta}|\boldsymbol{\pi})-q_0)/q_0$.
By summing the square of each of these contributions, we introduce the cost function $\mathcal{C}$ for our moment matching problem, i.e.,

\begin{equation}\label{eq:cost_function}
    \mathcal{C}(\boldsymbol{\theta}|\boldsymbol{\pi}) = \frac{1}{2}  \delta H(\boldsymbol{\theta}|\boldsymbol{\pi})^2+ \frac{1}{2} \delta Q_d(\boldsymbol{\theta}|\boldsymbol{\pi})^2 + \frac{1}{2}  \delta Q_0^2(\boldsymbol{\theta}|\boldsymbol{\pi}).
\end{equation}

As already observed, the cost function has multiple local minima. To explore the rich structure of minima, we adopt a greedy search optimization strategy. First, we generate a vector $\boldsymbol{\theta}_0$ so that $m_0=m(\boldsymbol{\theta}_0|\boldsymbol{\pi})=1/2 \ \text{max}(m)=1/2$. This condition ensures that the starting point of the optimization is far from the unbounded growth phase. In particular, it allows us to randomly generate an initial value of the interactions heterogeneity from a broad range $\sigma \sim \text{Uniform}(0,10)$ and to get the corresponding initial value of the demographic noise by mean of the relation $T_0 = \frac{(q_d-q_0)\sigma^2_0}{1-m_0} = 2(q_d-q_0)\sigma^2_0$. The other two parameters are randomly drawn, respectively, as $\mu \sim \text{Uniform}(-1,1)$ and $\log_{10} \lambda_0 \sim \text{Uniform}(-8,-3)$. The choice of the $\mu_0$ range is justified by the fact that we do not want to bias the interactions to be mutualistic or competitive. Since the unbounded growth phase emerges at $\mu_0=-1$ \cite{biroli2018marginally,bunin2017ecological}, one reasonable choice for the initial condition upper bound is $\mu_0=1$. Secondly, $\lambda_0$ is introduced as a regularizing term for the Langevin dynamics and can be considered small: in this way, we can bound its values as described above. Once the initialization is set, we optimize the cost function. To explore the largest set of solutions, we employ the Broyden–Fletcher–Goldfarb–Shanno algorithm provided by the scipy \cite{virtanen2020scipy} routine. Briefly, this method allows us to optimize scalar functions of multiple variables using a generalized secant method. Since $\mathcal{C}$ is flat almost everywhere except in the region where local minima are clustered, other methods tend to provide -- with the same initialization procedure -- results on the boundary of the optimization region, signaling poor convergence performance when tested for our problem. To explore a large subset of solutions and take the flatness of the cost function into account, we repeat the process $10^5$ times and bound the solutions into a region way larger than the initial conditions $\mu \in [-1,100]$, $\sigma \in [0,10]$, $T \in [10^{-4},10^{-2}]$ and $\lambda \in [10^{-9},10^{-1}]$. In the downstream analysis, we only retain the best 30 solutions, minimizing $\mathcal{C}$. At the end of the procedure, we obtain a set of parameters $\boldsymbol{\theta}=(\mu,\sigma,T,\lambda)$ that, if used in the self-consistent equations, are capable of satisfying them with mean relative error $\mathcal{E}=\delta H+\delta Q_d + \delta Q_0$ of order $\mathcal{E}\approx 10^{-2},10^{-3}$. As a consistency check, we report the value of $\lambda^* = 2 \times 10^{-6} \approx \min N_{data} = 9 \times 10^{-6}$ (constant for all of the top 30 solutions), which is slightly below the minimum species relative abundance of the data.

\subsection*{Cavity method for the species abundance distribution}\label{methods:cavity}

Another powerful technique rooted in disordered systems is the cavity method, which turns out to be particularly convenient for deriving the species abundance distribution at equilibrium.
Without demographic noise, the SAD in the single equilibrium phase is typically captured by a truncated Gaussian distribution \cite{yoshino2008rank, bunin2017ecological, altieri2019}. In the presence of noise and finite migration, the computation gets more involved but is still doable within the cavity approach \cite{mezard-montanari}.
The basic idea consists of adding a new species to the ecosystem and investigating the resulting joint probability distribution of the typical species. In the thermodynamic limit, the difference between a system composed of $S$ species and the corresponding one with $S+1$ species is negligible. Therefore, one can write:
\begin{equation}
P_{S+1}(\lbrace N_i \rbrace, N_c)\propto P_S(\lbrace N_i \rbrace) \frac{1}{N_c^{1-\lambda/T}} \exp \left[ \frac{1}{T} N_c \left( K-\frac{N_c}{2}-\sum_{j \neq i} \alpha_{c j} N_j \right)\right] \ .
\end{equation}
By gathering all relevant information about the so-called \emph{cavity field}, $h_c=\sum_{j} \alpha_{c j} N_j$, and the higher-order correlation term,
we obtain the field distribution, which is defined by the two moments
\begin{equation}
       \overline{ \tilde{h} }=\sum_j \overline{\alpha_{cj}} \overline{\langle N_j \rangle}=\mu h \hspace{0.3cm} \ , \hspace{0.25cm} \overline{ \tilde{h}^2} =\sum_{j,k} \overline{\alpha_{c i} \alpha_{c j}} \overline{  \langle N_i \rangle \langle N_j \rangle } =\sigma^2 q_0 \ .   \label{SP_methods}
\end{equation}
For compactness, we skip all technical details at this stage. Proceeding step-by-step -- the full derivation can nevertheless be found in the Appendix -- we end up writing the expression for the marginal probability distribution:
\begin{equation}
\begin{split}
P_{S+1}(N)& \simeq P_{S}(N)=
\int \mathcal{D}\zeta \frac{1}{\mathcal{Z}(\zeta)} N^{\beta \lambda -1 }  \exp \Biggl \lbrace  -\frac{\beta}{2}\left[ m N^2 -2 \zeta N \right] \Biggr \rbrace 
\label{marginal_p}
\end{split}
\end{equation}
where $N_c$ has been replaced by $N$ denoting the typical species abundance, $m=\left[1-\sigma^2\beta(q_d-q_0)\right]$ denotes the \emph{mass} term borrowing field-theory terminology, and $\zeta$ is an auxiliary Gaussian variable. 

\vspace{1cm}


\appendix

\label{first:app}

\section*{Appendices}

\subsection*{S1. Mapping between MacArthur and Lotka-Volterra Models}
\vspace{0.3cm}

We aim to investigate the mapping between the Lokta-Volterra model and the consumer-resource (CR) or MacArthur (MA) dynamics, as analyzed in \cite{tikhonov2017, altieri2019} for a high-dimensional version. We thus consider Eq. (1) of the main text, which describes the evolution of $S$ randomly interacting species. The self-interaction term parameters are the growth rate $r_i$, and the carrying capacity $K_i$, whereas all the inter-species interaction parameters are collected into the matrix $\boldsymbol{\alpha}$. For this argument, we will consider a simplified model version, where stochasticity (a.k.a. effective temperature) and immigration rate are set to zero, leading to a deterministic version, as in \cite{biroli2018marginally}. This model describes the dynamics of the relative species abundances $N_i$ -- previously normalized with respect to the total number of individuals in the pool -- according to:
\begin{equation}
    \frac{\dot{N}_i}{N_i }=\frac{r_i}{K_i}( K_i - N_i-\sum_{j \neq i } \alpha_{i,j}N_j)= r_i - \frac{r_i}{K_i} N_i -\sum_{j \neq i } \frac{r_i}{K_i} \alpha_{i,j}N_j = \rho K_i - \rho N_i -\sum_{j \neq i } \rho \alpha_{i,j}N_j  \ ,
\end{equation}
where we have introduced $\frac{r_i}{K_i}=\rho_i=\rho$, the proportionality constant between growth rate and carrying capacity. Our goal is to re-derive here a generalized Lotka-Volterra (gLV) model with symmetric interactions and therefore to justify the assumption adopted in the main text. To do that, we introduce the MacArthur model, which describes how $S$ species compete to grab $M$ available resources. For more details, see also \cite{wu2021understanding}. We thus consider the following set of coupled ODEs, describing the joint dynamics of resources abundances $R_{\alpha}$ and species abundances $N_i$:

\begin{equation}
\begin{cases}
    \dot{N}_i = N_i \ ( \sum_{\beta} C_{i,\beta} R_{\beta} - m_i) \\
    \dot{R}_{\alpha} = f_{\alpha}(R_{\alpha}) - R_{\alpha} \sum_{j} C^{T}_{\alpha,j} N_j  \ ,
\end{cases}
\end{equation}

where $f_{\alpha}=R_{\alpha}(\Tilde{K}_{\alpha}-R_\alpha)$ describes resources logistic regulation in the absence of consumer species. The parameters here are the death rate of each species $m_i$, the carrying capacity of each resource $\Tilde{K}_{\alpha}$, and the uptake matrix $\boldsymbol{C}$, which describes the consumption or production versus microbial growth inhibition of the resource $\beta$ per unit of time by the species $i$. Each row of the matrix can be considered an $M$-dimensional vector, $\Vec{C}_i$, which we will refer to as the uptake profile of the species $i$. The key point is that such vectors might have negative entries, as it happens in the case of microbial growth inhibition.
If the dynamics of resources is much faster than the one of species, we can consider the associated equation in the stationary state. With this approximation, provided that the resources abundance is larger than zero ($R_{\alpha}>0$), we can evaluate the stationary value of the resource abundances 

\begin{equation}
    \Tilde{K}_{\alpha}-R_\alpha - \sum_{j} C_{j,\alpha} N_j \longrightarrow R_\alpha = \Tilde{K}_{\alpha} - \sum_{j} C_{j,\alpha} N_j \ .
\end{equation}
By plugging the obtained  resource abundance in the equation for species dynamics, we precisely recover the \emph{Generalized Lotka-Volterra} model

\begin{align}\label{reduced_ma:eq}
    \frac{\dot{N}_i}{N_i } & = \sum_{\beta} \left[ C_{i,\beta} \left( \Tilde{K}_{\beta} - \sum_{j} C_{j,\beta} N_j  \right) \right] -m_i   =  \left(\sum_{\beta} C_{i,\beta} \Tilde{K}_{\beta}-m_i \right) - \sum_{\beta} \left(\sum_{j} C_{i,\beta} C^{T}_{\beta,j}\right) N_j = \\
    & =  r^{MA}_i - \left(\sum_{\beta} C_{i,\beta} C^{T}_{\beta,i} \right) N_i -  \sum_{j \neq i} \left(\sum_{\beta} C_{i,\beta} C^{T}_{\beta,j}\right) N_j  =  r^{MA}_i - A_{i,i} N_i - \sum_{j \neq i} A_{i,j} N_j  \ .\\
\end{align}
Note that, between the second and third passage, we decomposed the sum over $j$ in the diagonal and off-diagonal terms, respectively $j=i$ and $j \neq i$, then swapped the summation over resources and species. Finally, we have introduced the quantities
\begin{equation}
    r^{MA}_i = \sum_{\beta} C_{i,\beta} \Tilde{K}_{\beta}-m_i, \quad
    A_{i,j} = \sum_{\beta} C_{i,\beta} C^{T}_{\beta,j} \ .
\end{equation}
Since the interactions are expressed as a scalar product between uptake profiles, the derived equations correspond to a gLV model with symmetric interactions.

\vspace{0.6cm}

\subsection*{S2. Replica formalism for the disordered Generalized Lotka-Volterra model}
\label{second:app}
\vspace{0.3cm}

To obtain a full characterization of the different emergent phases in the disordered Generalized Lotka-Volterra model, we can take advantage of the replica method. 
Using the replica identity \cite{mezard1987spin, altieri2024introduction}, i.e. $\overline{\ln{Z}}=\lim \limits_{n \rightarrow 0} \frac{\overline{{Z^n}}-1}{n} $ -- which has been rigorously proven in some specific instances \cite{guerra2002, panchenko2007, dia2016} -- the disordered average is directly computed on the replicated partition function, for $n$ replicas of the same sample. The index $n$ is initially treated as an integer, however relying on the assumption that the analytical continuation $n \rightarrow 0$ exists.

\begin{equation}
    \overline{Z^n} = \int \prod_{i <j} d \alpha_{ij} \exp \left[-\sum_{i <j } \frac{(\alpha_{ij} - \mu /S)^2}{2 \sigma^2/S} \right] \int \prod_{a=1}^{n} \prod_{i } d N_i^{a} \exp\left[-\beta H(\lbrace N_i^{a} \rbrace) \right] \ ,
\end{equation}
where $\beta=1/T$ is the inverse of the demographic noise amplitude. The overline $\overline \cdot$ in the partition function denotes the average over the disorder, namely over the Gaussian variables $\alpha_{ij}$ with finite mean $\mu/S$ and variance $\sigma^2/S$. The index $a$ takes the number of replicas into account, with $a=1,...,n$.

The analytical treatment requires the introduction of the overlap matrix, $Q_{ab}$ -- whose diagonal value is $Q_{aa}$ -- $
Q_{ab}=\frac{1}{S} \sum \limits_{i=1}^{S}  N_i^ a N_i^b $, which can be easily parametrized in the replica space, as well as the average abundance $h_a=\frac{1}{S} \sum \limits_{i=1}^{S}  N_i^a$. For the latter, we will consider the uniformity condition, with $h_a=h$, $\forall a$.
We can thus re-express the free energy as a function of the aforementioned order parameters
\begin{equation}
 F= -\frac{1}{\beta n}\ln \int \prod \limits_{a,(a<b)} d Q_{ab}d Q_{aa} d h_{a} \; e^{S \mathcal{A}(Q_{ab}, Q_{aa},h_a) }
 \label{freeenergy_RS}
\end{equation}
which remains, at this level, as general as possible. The action $\mathcal{A}$ reads
\begin{equation}
\begin{split}
& \mathcal{A}(Q_{ab},Q_{aa},h_a)=  -\rho^2 \sigma^2 \beta^2 \sum \limits_{a<b}\frac{Q_{ab}^2}{2}-\rho^2 \sigma^2 \beta^2 \sum \limits_a \frac{Q_{aa}^2}{4}+\rho \mu \beta \sum_a \frac{h_a^2}{2}+\frac{1}{S}\sum \limits_i \ln Z_i \ ,
\end{split}
\label{action_replicas}
\end{equation}
to be eventually evaluated by the Laplace method, or saddle-point approximation, in the large $S$ limit. 

By resorting to the replica-symmetric (RS) ansatz -- according to which the overlap matrix is parametrized by two values only, $q_d$ and $q_0$, the diagonal value or self-overlap, and the off-diagonal value, respectively -- the last piece in Eq. (\ref{action_replicas}) can be expressed as a function of an effective Hamiltonian $\mathcal{H}_\text{eff}(\lbrace{N_i^a \rbrace})$. It essentially embeds the contribution of the quadratic Lotka-Volterra potential and the additional Lagrange multipliers enforcing the above expressions for $Q_{ab}, Q_{aa}$ and $h_a$.
\begin{equation}
\mathcal{H}_\text{eff}(\lbrace{N_i^a \rbrace})=
-\frac{\rho^2\sigma^2 \beta}{2} (q_d-q_0) \sum_a (N_i^a)^2 -\frac{\rho^2\sigma^2 \beta}{2} q_0 \left(\sum_a N_i^a\right)^2 +\sum_a \rho \mu h N_i^a +\sum_a V_i(N_i^a) +(T-\lambda) \sum_a \ln(N_i^a)
\end{equation}
The free energy is thus re-written as $F=-\lim \limits_{n \rightarrow 0} \ln{Z^{n}}/(\beta n)=-\lim \limits_{n \rightarrow 0} \mathcal{A}(q_d, q_0,h)/(\beta n)$.
However, integrating the quenched disorder out implies a further complication: replica indices turn out to be coupled, as is evident in the second term above. The next step will then require the introduction of an auxiliary Gaussian variable $z$, with zero mean and unit variance so that:
\begin{equation}
    Z_i= \int_{-\infty}^{+\infty} \frac{d z_i}{\sqrt{2 \pi}} e^{-z_i^2/2} \int \prod \limits_{a=1}^{n} d N_i^a e^{-\beta \sum \limits_a \mathcal{H}_\text{RS}(N_i^a; q_d, q_0, h, z_i) } \ ,
\end{equation}
with the associated replica symmetric Hamiltonian:
\begin{equation}
\begin{split}
    \mathcal{H}_\text{RS}(N_i^a,z_i)=&-\rho^2 \sigma^2 \beta (q_d-q_0) \frac{({N_i}^{a})^2}{2} +(\rho \mu h - z_i \rho \sqrt{q_0} \sigma){N_i}^{a} +V_i({N_i}^{a})+(T-\lambda) \ln{{N_i}^{a}}=\\
     =& \frac{({N_i^a})^{2}}{2}\left[\rho -\rho^2 \sigma^2 \beta(q_d-q_0)\right] + \left(\rho \mu h -z_i \rho \sigma \sqrt{q_0} -\rho K_i \right){N_i}^{a}+ (T-\lambda) \ln{{N_i^{a}}}  \ .
     \label{Hamil_RS}
    \end{split}
\end{equation}
Note that, although in the computation above we have considered a species-dependent $K_i$ to be as general as possible, in our work we have assumed $K_i=K$ for all species.
The full expressions of the parameters $h$, $q_0$, and $q_d$, to be obtained by a saddle-point approximation of the RS free energy have been reported in the Methods. They can be solved iteratively as also explained in \cite{altieri2021properties, lorenzana2022well}.

\vspace{0.3cm}

\subsubsection*{Comment on compositional abundances}
\vspace{0.3cm}

For purely theoretical purposes, dealing with absolute or relative abundances makes little difference. However, for the metagenomics data available to us, it is essential to perform the analysis as a function of relative abundances.
Forcing the abundances to be normalized is formally equivalent to adding a global constraint in the Hamiltonian such that $\sum \limits_{i} N_i=S$ through the Lagrange multiplier $\gamma$.
Accordingly, the original partition function can be recast as
\begin{equation}
    \overline{Z} = \int \prod_{i <j} d \alpha_{ij} \exp \left[-\sum_{i <j } \frac{(\alpha_{ij} - \mu /S)^2}{2 \sigma^2/S} \right] \int \prod_{i } d N_i \int_{-i \infty}^{i \infty} d \gamma \; \exp\left[-\beta H(\lbrace N_i \rbrace) -\gamma \sum_i(N_i-1) \right] \ .
\end{equation}
The procedure explained above remains exactly the same along with the introduction of $n$ replicas of the reference system.
Optimizing over $Q_{ab}$, $H_a$, and the parameter $\gamma_a$ things get easier within the RS approximation. In fact, $\gamma_a=\gamma$, satisfying a uniformity condition for all replicas.
The only visible difference is in the linear term of (\ref{Hamil_RS}) which now incorporates the explicit dependence on the multiplier $\gamma$ leading to
\begin{equation}
 \mathcal{H}_\text{RS}(N_i^a,z_i)=\frac{({N_i^a})^{2}}{2}\left[\rho -\rho^2 \sigma^2 \beta(q_d-q_0)\right] + \left(\rho \mu h -z_i \rho \sigma \sqrt{q_0} -\rho K_i  -\gamma \right){N_i}^{a}+ (T-\lambda) \ln{{N_i^{a}}}  \ .
\end{equation}
We have already discussed in the text how the $\mu$ parameter does not play a significant role in determining the different phases, confirming prior results by Bunin and collaborators \cite{bunin2017ecological, lorenzana2022well}. Once more, we can safely claim that the actual differences and emerging data clustering between healthy and sick samples are driven by the noise amplitude $T$ and the interaction heterogeneity $\sigma$.

\vspace{0.6cm}

\subsection*{S3. Stability analysis: Hessian matrix and zero modes}
\vspace{0.3cm}

To properly understand the stability of the single equilibrium phase and highlight possibly emergent multiple attractor regimes, we need to perform a stability computation. Computing the stability against external perturbations requires first the definition of the replicated partition function:
\begin{equation}
\begin{split}
Z_i& =  \int \prod_a d N_i^a  \exp \biggl [ \frac{\beta^2 \rho^2 \sigma^2}{2} \sum \limits_{a <b} Q_{ab} N_i^a N_i^b + \beta^2 \rho^2 \sigma^2 \sum \limits_a (N_i^a)^2 \frac{Q_{aa}}{2}  -\rho \beta \mu \sum \limits_a N_i^a h^a -\beta \sum_a V_i(N_i^a) +\\
&-\beta\sum_a (T-\lambda) \ln {N_i^a} \biggr ] \ ,
 \end{split} 
\end{equation}
then the study the harmonic fluctuation of the free energy around the RS solution. For the latter, the Hessian matrix of the action $\mathcal{A}$ is needed
\begin{equation}
\begin{split}
   \mathcal{M}_{abcd} & \equiv  -\frac{\partial^2 \mathcal{A}}{\partial Q_{ab}\partial Q_{cd}}=\beta^2 \rho^2 \sigma^2 \left[\delta_{(ab),(cd)}-(\beta^2 \rho^2 \sigma^2) \overline{\langle N^a N^b, N^c N^d \rangle_c}\right]
   \label{Mass}
   \end{split}
\end{equation}
where the subscript $\langle \cdot \rangle _c$ denotes the connected part of the correlation function.
Leveraging well-known techniques in field theory and disordered systems \cite{dedominicis2006}, we decompose the Hessian matrix [\ref{Mass}] as a function of three different correlators among which the one living in the replicon subspace \cite{dedominicis2006, altieri2016composite} provides the leading contribution:
\begin{equation}
\mathcal{R}=   (\beta \rho \sigma)^2  \biggl [ 1-(\beta \rho \sigma)^2 \left( \tilde{M}_{ab,ab}-2\tilde{M}_{ab,ac}+\tilde{M}_{ab,cd} \right) \biggr ] \ .
\end{equation}
Three different elements appear in the computation depending on the constraint/equality condition of their replica indices 
\begin{equation}
\begin{split}
& \tilde{M}_{ab,ab}-2\tilde{M}_{ab,ac}+\tilde{M}_{ab,cd} =   \left[ \overline{\langle (N^a)^2(N^b)^2\rangle} -2 \overline{\langle (N^a)^2 N^b N^c \rangle }+\overline{\langle N^a N^b N^c N^d \rangle }\right] \ .
    \end{split}
\end{equation}
Sticking to a single-equilibrium regime, namely in the RS approximation, the expression for the replicon eigenvalue can be further simplified 
\begin{equation}
\mathcal{R}=(\beta \rho \sigma)^2 \left[ 1-(\beta \rho \sigma)^2 \overline{ \left(\langle N^2 \rangle - \langle N \rangle^2 \right)^2} \right] \ ,
\end{equation}
which precisely corresponds to Eq. (3) of the main text.
A strictly positive or a vanishing value of the replicon are associated with a stable or marginally stable phase, respectively. In the second case, the RS ansatz is no longer valid and requires the use of a one-step or multiple-step replica-symmetry breaking.

\vspace{0.6cm}

\subsection*{S4. Cavity argument for the species abundance distribution}\label{methods:cavity}
\vspace{0.3cm}

In the absence of demographic fluctuations and immigration, it is well-known that the species abundance distribution reflects a truncated Gaussian \cite{yoshino2008rank, bunin2017ecological, tikhonov2017, lorenzana2022well}.
Albeit the computation gets more complicated in the presence of demographic noise, we can nevertheless employ the cavity method \cite{mezard-montanari, zamponi2010, altieri2024introduction} to obtain the single-species marginal probability distribution.
Based on a cavity argument, we pretend to add a new species to the ecosystem and eventually investigate the resulting joint probability distribution. In the thermodynamic limit, for $S \gg 1$, it reads
\begin{equation}
P_{S+1}(\lbrace N_i \rbrace, N_c)\propto P_S(\lbrace N_i \rbrace) \frac{1}{N_c^{1-\lambda/T}} \exp \left[ \frac{1}{T} N_c \left( K-\frac{N_c}{2}-\sum_{j \neq i} \alpha_{c j} N_j \right)\right]
\end{equation}
where, for the sake of simplicity, we have assumed the carrying capacities to be species-independent, i.e. $K_i=K$. 

The last piece in parentheses denotes the \emph{cavity field}, $h_c=\sum_{j} \alpha_{c j} N_j$ which, as long as the single equilibrium phase is concerned, can be assumed to be a Gaussian random variable. The probability distribution is formally factorized and the only coupling with the new species occurs via $h_c$:
\begin{equation}
   P_{S+1}(N_c) \propto  \int_{0}^{\infty} \prod_i d N_i P_S(\lbrace N_i\rbrace) \frac{1}{N_c ^{1-\lambda/T}} \exp \left[\frac{1}{T}N_c \left(K-\frac{N_c}{2}-\sum_{j \neq i} \alpha_{c j}N_j \right) \right] 
 \end{equation}

The probability distribution conditioned to $h_c$ then reads:
 \begin{equation}
    P_{S+1}(N_c \vert h_c) \propto 
    \frac{1}{N_c^{1-\lambda/T}} \exp \left[\frac{1}{T}N_c \left(K-\frac{N_c}{2}-h_c \right) \right] P(h_c) \ ,
    \end{equation}
where the thermal and the disordered averages of the cavity field, $h_c$, must eventually be evaluated.   Furthermore, to go beyond a naïve mean-field approximation, we need to subtract the background effect exerted by all other species on the picked one through an \emph{Onsager reaction term}. 
    
   \begin{equation}
   \begin{split} 
   P_{S+1}(N_c) =& \frac{1}{Z} \int_{-\infty}^{+\infty} \frac{dh_c}{\sqrt{2\pi \text{var}[h_c]}} \frac{1}{N_c^{1-\lambda/T}} \exp \left[ \frac{1}{T}N_c \left(K-\frac{N_c}{2}-h_c \right) \right] \exp \left(\frac{-(h_c-\tilde{h})^2}{2\text{var}[h_c]} \right) =\\
   =&\frac{1}{Z} \frac{1}{N_c^{1-\lambda/T}} \exp \Biggl \lbrace -\frac{N_c}{2T^2}\left[ N_c (T-\text{var}[h_c])-2K T+2 \tilde{h} T \right] \Biggr \rbrace
   \end{split}
   \label{P_cavity}
\end{equation}
In the above expression, $\tilde{h} = \langle h_c \rangle =\sum_j \alpha_{c j} \langle N_j\rangle$ and $\text{var}[h_c]=\langle h_c^2 \rangle -\langle h_c\rangle^2$ stand respectively for the first moment and the variance of the field with respect to the thermal average, in the thermodynamic limit.
The normalization, $Z$, can formally be expressed as a combination of error and hypergeometric functions as follows
\begin{equation}
\begin{split}
Z=& \int_{0}^{\infty} \frac{d N_c}{N_c^{1-\lambda/T}} \exp \Biggl \lbrace -\frac{N_c}{2T^2}\left[ N_c (T-\text{var}[h_c])-2K T+2 \tilde{h} T \right] \Biggr \rbrace=\\
=& 2^{-1+\frac{\lambda}{2T}} 
\left(\frac{T-\text{var}[h_c]}{T^2}\right)^{-\frac{\lambda}{2T}}\Biggl ( \text{Gamma}\left[\frac{\lambda}{2T} \right] \text{HypergeometricF1}\left[\frac{\lambda}{2T},\frac{1}{2},\frac{(K-\tilde{h})^2}{2(T- \text{var}[h_c])} \right]+\\
+& \frac{\sqrt{2}(-\tilde{h}+K)}{\sqrt{T-\text{var}[h_c]}}\; \text{Gamma} \left[ \frac{T+\lambda}{2T}\right]\text{HypergeometricF1}\left[ \frac{T+\lambda}{2T},\frac{3}{2},\frac{(K-\tilde{h})^2}{2(T- \text{var}[h_c])}  \right] \Biggr )  \ .
\label{normalization}\end{split}
\end{equation}
The next step requires the computation of the disordered average. By doing so, we end up with the distribution of the second field, which is defined in terms of the two moments:
\begin{equation}
        \overline{ \tilde{h} }=\sum_j \overline{\alpha_{cj}} \overline{\langle N_j \rangle}=\mu h \hspace{0.1cm} \ , \hspace{0.5cm} \overline{ \tilde{h}^2} =\sum_{j,k} \overline{\alpha_{c i} \alpha_{c j}} \overline{  \langle N_i \rangle \langle N_j \rangle } =\sigma^2 q_0 \ .
\end{equation}
A similar derivation was obtained in the case of a generalized Lotka-Volterra model with noise to be compared with numerical distributions in different regimes \cite{wu2021understanding}. The resulting species abundance distribution typically exhibits a divergence at $N = 0$ and a secondary maximum, therefore closer to what is referred to as \emph{niche scenario}.

Then, by introducing a rescaled random variable $\zeta$ and integrating over it, the marginal probability distribution becomes
\begin{equation}\label{eq:methods_cavity_sad}
\begin{split}
P_{S+1}(N_c)&=\int_{-\infty}^{\infty} \frac{d \zeta}{\sigma \sqrt{2\pi  q_0}}e^{-\frac{\left[\zeta -(K-\mu h \right)]^2}{2 \sigma^2 q_0}} \frac{1}{\mathcal{Z}(\zeta) N_c^{1-\lambda/T}} \exp \Biggl \lbrace  -\frac{\beta}{2}\left[ N_c^2(1-\sigma^2\beta (q_d-q_0))-2N_c \zeta \right] \Biggr \rbrace \\
& \rightarrow \int \mathcal{D}\zeta \frac{1}{\mathcal{Z}(\zeta)} N^{\beta \lambda -1 }  \exp \Biggl \lbrace  -\frac{\beta}{2}\left[ m N^2 -2 \zeta N \right] \Biggr \rbrace 
\end{split}
\end{equation}
where $N_c$ is just replaced by $N$ denoting the typical species abundance and $T \rightarrow 1/\beta$. The expression for the normalization is roughly similar to the one in Eq. (\ref{P_cavity}) with the only difference being that the mean and the variance of the cavity field must be replaced by the quenched average of $\tilde{h}$ and $\sigma^ 2 (q_d-q_0)$, respectively.

\vspace{0.6cm}

\subsection*{S5. Estimation of Neutral and Niche processes}
\vspace{0.3cm}

Following the results of the cavity method in Section \ref{methods:cavity}, we can describe the species abundance distribution of the system (see Eq. (\ref{eq:methods_cavity_sad})) via the quantities $\nu=\beta\lambda>0$, the quadratic coefficient $m=1-\sigma^2 \beta (q_d-q_0)$ and the disorder variable $\zeta = K-\mu h + \sqrt{q_0}\sigma z$, where $z$ is a standardized normal variable.

As we specified in the main text, we refer to the quadratic coefficient $m$ as a mass. When the mass becomes negative, the self-consistent equations (\ref{eq:sp}) in the Methods section are no longer well-defined, and the order parameters diverge. This signals a transition to an unbounded growth regime.

So far, the obtained SAD is flexible enough to accommodate different qualitative shapes frequently appearing in theoretical ecology \cite{azaele2016statistical}. When $\nu>1$, the migration strength turns out to be stronger than demographic fluctuations and the distribution resembles a modal log-normal. On the other hand, when $0<\nu<1$ demographic noise overcomes immigration, and a low-abundance (integrable) divergence appears. In our case, $\nu \approx 10^{-3}$, so demographic noise is much stronger than immigration. 

When the divergence in zero appears, two further shapes are possible. If there exists a secondary extremum for the species abundance distribution, then there is an coexistence of  niche (driven by species interactions) and neutral (driven by demographic noise) effects. Conversely, if no secondary extrema exist, birth and death effects are the key drivers of the dynamics. In our interpretation of the model, the species abundance distribution is the stationary distribution of a \emph{single} realization of the Lotka-Volterra dynamics. Given a set of parameters $\boldsymbol{\theta}=(\mu,\sigma,\beta,\lambda)$, it is possible to evaluate how many realizations of the Lotka-Volterra dynamics will display a secondary extremum. Taking the form of a fraction, we introduce a quantity $\psi$ that captures the chance of displaying such an extremum and measures the relative contribution between selective and neutral effects. Namely, we can evaluate the probability that $\mathcal{H}_{\text{RS}}$ has a minimum different from zero:

\begin{equation}\label{eq:RS_extrema}
     \frac{\partial p(N|\zeta)}{\partial N} = - \beta \frac{\partial \mathcal{H}_{\text{RS}}}{\partial N} p(N|\zeta) = 0 \rightarrow \frac{\partial \mathcal{H}_{\text{RS}}}{\partial N} = \frac{1-\nu}{\beta N} + m N - \zeta  = 0 \ .
\end{equation}

Being an algebraic equation of degree two, it has two solutions $N_{\pm}^* = \frac{\zeta \pm \sqrt{\zeta^2 - 4 m (1-\nu)/\beta}}{2m}$. Introducing $\zeta^*=\sqrt{\frac{4(1-\nu)m}{\beta}}$, the condition for the local extrema to appear if $\zeta^2 > \frac{4 m (1-\nu)}{\beta } = (\zeta^*)^2 > 0$. This probability amounts to

\begin{equation}
    \psi=\mathbb{P}[\zeta^2 > (\zeta^*)^2] = \mathbb{P} \left [\zeta > \zeta^* \right ] + \mathbb{P} \left [\zeta < - \zeta^* \right ] = \frac{1}{2}  \text{Erfc} \left ( \frac{ \zeta^*+\overline{\zeta }}{\sqrt{2} \sigma_{\zeta}} \right ) + \frac{1}{2} \text{Erfc} \left ( \frac{\zeta^*-\overline{\zeta }}{\sqrt{2} \sigma_{\zeta}} \right ) \ ,
\end{equation}
where $\overline{\zeta}=K-\mu h$, $\sigma_{\zeta}=\sqrt{q_0}\sigma$ are the moments of the variable tracking the different disorder realizations. When the quantity $\psi$ is of order $1$, niche and neutral contributions to the dynamics can be considered comparable. In principle, we should check which of the two $\pm$ solutions is actually a maximum.  The convexity of the SAD is easy to evaluate:

\begin{equation}
\begin{split}
    \frac{\partial^2 p(N|\zeta)}{\partial N^2} & = -\beta  \left ( \left ( \frac{\partial^2 \mathcal{H}_{\text{RS}}}{\partial N^2} - \beta \left ( \frac{\partial \mathcal{H}_{\text{RS}}}{\partial N} \right )^2 \right ) p(N|\zeta)  \right |_{N=N_{\pm}^*} = \\
    &= \left ( \frac{\partial^2 \mathcal{H}_{\text{RS}}}{\partial N^2} - \beta \left ( \frac{\partial \mathcal{H}_{\text{RS}}}{\partial N} \right ) \right |_{N^*_{\pm}} = 2m(1 + \frac{1}{y} \left ( 1 \mp \sqrt{1-y} \right ) < 0
\end{split}
\end{equation}

 where $y=(\frac{\zeta^*}{\zeta})^2$. From the first inequality, straightforward algebraic manipulations confirm that $N^*_{+}$ is the stable solution. Intuitively, since $P(N|\zeta)$ has a divergence in zero, any secondary maximum $N^*_+$ -- if it exists -- must be greater than the minimum $N^*_-$.

\bibliographystyle{unsrtnat}
\bibliography{biblio2} 

\providecommand{\noopsort}[1]{}\providecommand{\singleletter}[1]{#1}%
\begin{thebibliography}{69}
\providecommand{\natexlab}[1]{#1}
\providecommand{\url}[1]{\texttt{#1}}
\expandafter\ifx\csname urlstyle\endcsname\relax
  \providecommand{\doi}[1]{doi: #1}\else
  \providecommand{\doi}{doi: \begingroup \urlstyle{rm}\Url}\fi

\bibitem[Levy et~al.(2016)Levy, Thaiss, and Elinav]{levy2016metabolites}
Maayan Levy, Christoph~A Thaiss, and Eran Elinav.
\newblock Metabolites: messengers between the microbiota and the immune system.
\newblock \emph{Genes \& development}, 30\penalty0 (14):\penalty0 1589--1597,
  2016.

\bibitem[Barroso-Batista et~al.(2015)Barroso-Batista, Demengeot, and
  Gordo]{barroso2015adaptive}
Jo{\~a}o Barroso-Batista, Jocelyne Demengeot, and Isabel Gordo.
\newblock Adaptive immunity increases the pace and predictability of
  evolutionary change in commensal gut bacteria.
\newblock \emph{Nature communications}, 6\penalty0 (1):\penalty0 8945, 2015.

\bibitem[Barreto and Gordo(2023)]{barreto2023}
Hugo~C Barreto and Isabel Gordo.
\newblock Intrahost evolution of the gut microbiota.
\newblock \emph{Nature Reviews Microbiology}, 21\penalty0 (9):\penalty0
  590--603, 2023.

\bibitem[Lloyd-Price et~al.(2019)Lloyd-Price, Arze, Ananthakrishnan, Schirmer,
  Avila-Pacheco, Poon, Andrews, Ajami, Bonham, Brislawn,
  et~al.]{lloyd2019multi}
Jason Lloyd-Price, Cesar Arze, Ashwin~N Ananthakrishnan, Melanie Schirmer,
  Julian Avila-Pacheco, Tiffany~W Poon, Elizabeth Andrews, Nadim~J Ajami,
  Kevin~S Bonham, Colin~J Brislawn, et~al.
\newblock Multi-omics of the gut microbial ecosystem in inflammatory bowel
  diseases.
\newblock \emph{Nature}, 569\penalty0 (7758):\penalty0 655--662, 2019.

\bibitem[Das and Nair(2019)]{das2019homeostasis}
Bhabatosh Das and G~Balakrish Nair.
\newblock Homeostasis and dysbiosis of the gut microbiome in health and
  disease.
\newblock \emph{Journal of biosciences}, 44:\penalty0 1--8, 2019.

\bibitem[Nishida et~al.(2018)Nishida, Inoue, Inatomi, Bamba, Naito, and
  Andoh]{nishida2018gut}
Atsushi Nishida, Ryo Inoue, Osamu Inatomi, Shigeki Bamba, Yuji Naito, and Akira
  Andoh.
\newblock Gut microbiota in the pathogenesis of inflammatory bowel disease.
\newblock \emph{Clinical journal of gastroenterology}, 11:\penalty0 1--10,
  2018.

\bibitem[Zeng et~al.(2015)Zeng, Sukumaran, Wu, and Rodrigo]{zeng2015neutral}
Qinglong Zeng, Jeet Sukumaran, Steven Wu, and Allen Rodrigo.
\newblock Neutral models of microbiome evolution.
\newblock \emph{PLoS computational biology}, 11\penalty0 (7):\penalty0
  e1004365, 2015.

\bibitem[Venkataraman et~al.(2015)Venkataraman, Bassis, Beck, Young, Curtis,
  Huffnagle, and Schmidt]{venkataraman2015application}
Arvind Venkataraman, Christine~M Bassis, James~M Beck, Vincent~B Young,
  Jeffrey~L Curtis, Gary~B Huffnagle, and Thomas~M Schmidt.
\newblock Application of a neutral community model to assess structuring of the
  human lung microbiome.
\newblock \emph{MBio}, 6\penalty0 (1):\penalty0 10--1128, 2015.

\bibitem[Sala et~al.(2016)Sala, Vitali, Giampieri, do~Valle, Remondini,
  Garagnani, Bersanelli, Mosca, Milanesi, and Castellani]{sala2016stochastic}
Claudia Sala, Silvia Vitali, Enrico Giampieri, {\`I}talo~Faria do~Valle, Daniel
  Remondini, Paolo Garagnani, Matteo Bersanelli, Ettore Mosca, Luciano
  Milanesi, and Gastone Castellani.
\newblock Stochastic neutral modelling of the gut microbiota’s relative
  species abundance from next generation sequencing data.
\newblock \emph{BMC bioinformatics}, 17\penalty0 (2):\penalty0 179--188, 2016.

\bibitem[Sieber et~al.(2019)Sieber, Pita, Weiland-Br{\"a}uer, Dirksen, Wang,
  Mortzfeld, Franzenburg, Schmitz, Baines, Fraune,
  et~al.]{sieber2019neutrality}
Michael Sieber, Luc{\'\i}a Pita, Nancy Weiland-Br{\"a}uer, Philipp Dirksen, Jun
  Wang, Benedikt Mortzfeld, S{\"o}ren Franzenburg, Ruth~A Schmitz, John~F
  Baines, Sebastian Fraune, et~al.
\newblock Neutrality in the metaorganism.
\newblock \emph{PLoS Biology}, 17\penalty0 (6):\penalty0 e3000298, 2019.

\bibitem[Descheemaeker and De~Buyl(2020)]{descheemaeker2020stochastic}
Lana Descheemaeker and Sophie De~Buyl.
\newblock Stochastic logistic models reproduce experimental time series of
  microbial communities.
\newblock \emph{Elife}, 9:\penalty0 e55650, 2020.

\bibitem[Grilli(2020)]{grilli2020macroecological}
Jacopo Grilli.
\newblock Macroecological laws describe variation and diversity in microbial
  communities.
\newblock \emph{Nature communications}, 11\penalty0 (1):\penalty0 4743, 2020.

\bibitem[Zaoli and Grilli(2022)]{zaoli2022stochastic}
Silvia Zaoli and Jacopo Grilli.
\newblock The stochastic logistic model with correlated carrying capacities
  reproduces beta-diversity metrics of microbial communities.
\newblock \emph{PLOS Computational Biology}, 18\penalty0 (4):\penalty0
  e1010043, 2022.

\bibitem[Azaele et~al.(2016)Azaele, Suweis, Grilli, Volkov, Banavar, and
  Maritan]{azaele2016statistical}
Sandro Azaele, Samir Suweis, Jacopo Grilli, Igor Volkov, Jayanth~R Banavar, and
  Amos Maritan.
\newblock Statistical mechanics of ecological systems: Neutral theory and
  beyond.
\newblock \emph{Reviews of Modern Physics}, 88\penalty0 (3):\penalty0 035003,
  2016.

\bibitem[Seppi et~al.(2023)Seppi, Pasqualini, Facchin, Savarino, and
  Suweis]{seppi2023emergent}
Marcello Seppi, Jacopo Pasqualini, Sonia Facchin, Edoardo~Vincenzo Savarino,
  and Samir Suweis.
\newblock Emergent functional organization of gut microbiomes in health and
  diseases.
\newblock \emph{Biomolecules}, 14\penalty0 (1):\penalty0 5, 2023.

\bibitem[Pasqualini et~al.(2024)Pasqualini, Facchin, Rinaldo, Maritan,
  Savarino, and Suweis]{pasqualini2024emergent}
Jacopo Pasqualini, Sonia Facchin, Andrea Rinaldo, Amos Maritan, Edoardo
  Savarino, and Samir Suweis.
\newblock Emergent ecological patterns and modelling of gut microbiomes in
  health and in disease.
\newblock \emph{PLOS Computational Biology}, 20\penalty0 (9):\penalty0
  e1012482, 2024.

\bibitem[Bashan et~al.(2016)Bashan, Gibson, Friedman, Carey, Weiss, Hohmann,
  and Liu]{bashan2016universality}
Amir Bashan, Travis~E Gibson, Jonathan Friedman, Vincent~J Carey, Scott~T
  Weiss, Elizabeth~L Hohmann, and Yang-Yu Liu.
\newblock Universality of human microbial dynamics.
\newblock \emph{Nature}, 534\penalty0 (7606):\penalty0 259--262, 2016.

\bibitem[Kehe et~al.(2021)Kehe, Ortiz, Kulesa, Gore, Blainey, and
  Friedman]{kehe2021positive}
Jared Kehe, Anthony Ortiz, Anthony Kulesa, Jeff Gore, Paul~C Blainey, and
  Jonathan Friedman.
\newblock Positive interactions are common among culturable bacteria.
\newblock \emph{Science advances}, 7\penalty0 (45):\penalty0 eabi7159, 2021.

\bibitem[Venturelli et~al.(2018)Venturelli, Carr, Fisher, Hsu, Lau, Bowen,
  Hromada, Northen, and Arkin]{venturelli2018deciphering}
Ophelia~S Venturelli, Alex~V Carr, Garth Fisher, Ryan~H Hsu, Rebecca Lau,
  Benjamin~P Bowen, Susan Hromada, Trent Northen, and Adam~P Arkin.
\newblock Deciphering microbial interactions in synthetic human gut microbiome
  communities.
\newblock \emph{Molecular systems biology}, 14\penalty0 (6):\penalty0 e8157,
  2018.

\bibitem[Faust and Raes(2012)]{faust2012microbial}
Karoline Faust and Jeroen Raes.
\newblock Microbial interactions: from networks to models.
\newblock \emph{Nature Reviews Microbiology}, 10\penalty0 (8):\penalty0
  538--550, 2012.

\bibitem[Xiao et~al.(2017)Xiao, Angulo, Friedman, Waldor, Weiss, and
  Liu]{xiao2017mapping}
Yandong Xiao, Marco~Tulio Angulo, Jonathan Friedman, Matthew~K Waldor, Scott~T
  Weiss, and Yang-Yu Liu.
\newblock Mapping the ecological networks of microbial communities.
\newblock \emph{Nature communications}, 8\penalty0 (1):\penalty0 2042, 2017.

\bibitem[Camacho-Mateu et~al.(2024)Camacho-Mateu, Lampo, Sireci, Mu{\~n}oz, and
  Cuesta]{camacho2024sparse}
Jos{\'e} Camacho-Mateu, Aniello Lampo, Matteo Sireci, Miguel~A Mu{\~n}oz, and
  Jos{\'e}~A Cuesta.
\newblock Sparse species interactions reproduce abundance correlation patterns
  in microbial communities.
\newblock \emph{Proceedings of the National Academy of Sciences}, 121\penalty0
  (5):\penalty0 e2309575121, 2024.

\bibitem[Faisal et~al.(2010)Faisal, Dondelinger, Husmeier, and
  Beale]{faisal2010inferring}
Ali Faisal, Frank Dondelinger, Dirk Husmeier, and Colin~M Beale.
\newblock Inferring species interaction networks from species abundance data: A
  comparative evaluation of various statistical and machine learning methods.
\newblock \emph{Ecological Informatics}, 5\penalty0 (6):\penalty0 451--464,
  2010.

\bibitem[Angulo et~al.(2017)Angulo, Moreno, Lippner, Barab{\'a}si, and
  Liu]{angulo2017fundamental}
Marco~Tulio Angulo, Jaime~A Moreno, Gabor Lippner, Albert-L{\'a}szl{\'o}
  Barab{\'a}si, and Yang-Yu Liu.
\newblock Fundamental limitations of network reconstruction from temporal data.
\newblock \emph{Journal of the Royal Society Interface}, 14\penalty0
  (127):\penalty0 20160966, 2017.

\bibitem[Tu et~al.(2019)Tu, Suweis, Grilli, Formentin, and
  Maritan]{tu2019reconciling}
Chengyi Tu, Samir Suweis, Jacopo Grilli, Marco Formentin, and Amos Maritan.
\newblock Reconciling cooperation, biodiversity and stability in complex
  ecological communities.
\newblock \emph{Scientific reports}, 9\penalty0 (1):\penalty0 5580, 2019.

\bibitem[Armitage and Jones(2019)]{armitage2019sample}
David~W Armitage and Stuart~E Jones.
\newblock How sample heterogeneity can obscure the signal of microbial
  interactions.
\newblock \emph{The ISME journal}, 13\penalty0 (11):\penalty0 2639--2646, 2019.

\bibitem[Holt(2020)]{holt2020some}
Robert~D Holt.
\newblock Some thoughts about the challenge of inferring ecological
  interactions from spatial data.
\newblock \emph{Biodiversity Informatics}, 15\penalty0 (1):\penalty0 61--66,
  2020.

\bibitem[Pacciani-Mori et~al.(2021)Pacciani-Mori, Suweis, Maritan, and
  Giometto]{pacciani2021constrained}
Leonardo Pacciani-Mori, Samir Suweis, Amos Maritan, and Andrea Giometto.
\newblock Constrained proteome allocation affects coexistence in models of
  competitive microbial communities.
\newblock \emph{The ISME Journal}, 15\penalty0 (5):\penalty0 1458--1477, 2021.

\bibitem[Sireci et~al.(2022)Sireci, Mu{\~n}oz, and
  Grilli]{sireci2022environmental}
Matteo Sireci, Miguel~A Mu{\~n}oz, and Jacopo Grilli.
\newblock Environmental fluctuations explain the universal decay of
  species-abundance correlations with phylogenetic distance.
\newblock \emph{bioRxiv}, pages 2022--07, 2022.

\bibitem[Tovo et~al.(2020)Tovo, Menzel, Krogh, Cosentino~Lagomarsino, and
  Suweis]{tovo2020taxonomic}
Anna Tovo, Peter Menzel, Anders Krogh, Marco Cosentino~Lagomarsino, and Samir
  Suweis.
\newblock Taxonomic classification method for metagenomics based on core
  protein families with core-kaiju.
\newblock \emph{Nucleic Acids Research}, 48\penalty0 (16):\penalty0 e93--e93,
  2020.

\bibitem[Weiss et~al.(2016)Weiss, Van~Treuren, Lozupone, Faust, Friedman, Deng,
  Xia, Xu, Ursell, Alm, et~al.]{weiss2016correlation}
Sophie Weiss, Will Van~Treuren, Catherine Lozupone, Karoline Faust, Jonathan
  Friedman, Ye~Deng, Li~Charlie Xia, Zhenjiang~Zech Xu, Luke Ursell, Eric~J
  Alm, et~al.
\newblock Correlation detection strategies in microbial data sets vary widely
  in sensitivity and precision.
\newblock \emph{The ISME journal}, 10\penalty0 (7):\penalty0 1669--1681, 2016.

\bibitem[Gloor et~al.(2017)Gloor, Macklaim, Pawlowsky-Glahn, and
  Egozcue]{gloor2017microbiome}
Gregory~B Gloor, Jean~M Macklaim, Vera Pawlowsky-Glahn, and Juan~J Egozcue.
\newblock Microbiome datasets are compositional: and this is not optional.
\newblock \emph{Frontiers in microbiology}, 8:\penalty0 2224, 2017.

\bibitem[Dohlman and Shen(2019)]{dohlman2019mapping}
Anders~B Dohlman and Xiling Shen.
\newblock Mapping the microbial interactome: Statistical and experimental
  approaches for microbiome network inference.
\newblock \emph{Experimental Biology and Medicine}, 244\penalty0 (6):\penalty0
  445--458, 2019.

\bibitem[May(1972)]{may1972will}
Robert~M May.
\newblock Will a large complex system be stable?
\newblock \emph{Nature}, 238:\penalty0 413--414, 1972.

\bibitem[Allesina and Tang(2012)]{allesina2012stability}
Stefano Allesina and Si~Tang.
\newblock Stability criteria for complex ecosystems.
\newblock \emph{Nature}, 483\penalty0 (7388):\penalty0 205--208, 2012.

\bibitem[Bunin(2017)]{bunin2017ecological}
Guy Bunin.
\newblock Ecological communities with lotka-volterra dynamics.
\newblock \emph{Physical Review E}, 95\penalty0 (4):\penalty0 042414, 2017.

\bibitem[Galla(2018)]{galla2018dynamically}
Tobias Galla.
\newblock Dynamically evolved community size and stability of random
  lotka-volterra ecosystems (a).
\newblock \emph{Europhysics Letters}, 123\penalty0 (4):\penalty0 48004, 2018.

\bibitem[Biroli et~al.(2018)Biroli, Bunin, and Cammarota]{biroli2018marginally}
Giulio Biroli, Guy Bunin, and Chiara Cammarota.
\newblock Marginally stable equilibria in critical ecosystems.
\newblock \emph{New Journal of Physics}, 20\penalty0 (8):\penalty0 083051,
  2018.

\bibitem[Altieri et~al.(2021)Altieri, Roy, Cammarota, and
  Biroli]{altieri2021properties}
Ada Altieri, Felix Roy, Chiara Cammarota, and Giulio Biroli.
\newblock Properties of equilibria and glassy phases of the random
  lotka-volterra model with demographic noise.
\newblock \emph{Physical Review Letters}, 126\penalty0 (25):\penalty0 258301,
  2021.

\bibitem[Lorenzana and Altieri(2022)]{lorenzana2022well}
Giulia~Garcia Lorenzana and Ada Altieri.
\newblock Well-mixed lotka-volterra model with random strongly competitive
  interactions.
\newblock \emph{Physical Review E}, 105\penalty0 (2):\penalty0 024307, 2022.

\bibitem[Barbier et~al.(2018)Barbier, Arnoldi, Bunin, and
  Loreau]{barbier2018generic}
Matthieu Barbier, Jean-Fran{\c{c}}ois Arnoldi, Guy Bunin, and Michel Loreau.
\newblock Generic assembly patterns in complex ecological communities.
\newblock \emph{Proceedings of the National Academy of Sciences}, 115\penalty0
  (9):\penalty0 2156--2161, 2018.

\bibitem[Hatton et~al.(2024)Hatton, Mazzarisi, Altieri, and
  Smerlak]{doi:10.1126/science.adg8488}
Ian~A. Hatton, Onofrio Mazzarisi, Ada Altieri, and Matteo Smerlak.
\newblock Diversity begets stability: Sublinear growth and competitive
  coexistence across ecosystems.
\newblock \emph{Science}, 383\penalty0 (6688):\penalty0 eadg8488, 2024.
\newblock \doi{10.1126/science.adg8488}.
\newblock URL \url{https://www.science.org/doi/abs/10.1126/science.adg8488}.

\bibitem[Hu et~al.(2022)Hu, Amor, Barbier, Bunin, and Gore]{hu2022emergent}
Jiliang Hu, Daniel~R Amor, Matthieu Barbier, Guy Bunin, and Jeff Gore.
\newblock Emergent phases of ecological diversity and dynamics mapped in
  microcosms.
\newblock \emph{Science}, 378\penalty0 (6615):\penalty0 85--89, 2022.

\bibitem[Altieri and Biroli(2022)]{Altieri2022}
Ada Altieri and Giulio Biroli.
\newblock Effects of intraspecific cooperative interactions in large
  ecosystems.
\newblock \emph{SciPost Physics}, 12\penalty0 (1):\penalty0 013, 2022.

\bibitem[Mallmin et~al.(2024)Mallmin, Traulsen, and
  De~Monte]{mallmin2024chaotic}
Emil Mallmin, Arne Traulsen, and Silvia De~Monte.
\newblock Chaotic turnover of rare and abundant species in a strongly
  interacting model community.
\newblock \emph{Proceedings of the National Academy of Sciences}, 121\penalty0
  (11):\penalty0 e2312822121, 2024.

\bibitem[Suweis et~al.(2024)Suweis, Ferraro, Grilletta, Azaele, and
  Maritan]{suweis2024generalized}
Samir Suweis, Francesco Ferraro, Christian Grilletta, Sandro Azaele, and Amos
  Maritan.
\newblock Generalized lotka-volterra systems with time correlated stochastic
  interactions.
\newblock \emph{Physical Review Letters}, 133\penalty0 (16):\penalty0 167101,
  2024.

\bibitem[Mezard and Montanari(2009)]{mezard-montanari}
Marc Mezard and Andrea Montanari.
\newblock \emph{Information, physics, and computation}.
\newblock Oxford University Press, 2009.

\bibitem[Altieri and Baity-Jesi(2024)]{altieri2024introduction}
Ada Altieri and Marco Baity-Jesi.
\newblock An introduction to the theory of spin glasses.
\newblock \emph{Encyclopedia of Condensed Matter Physics}, pages 361--370,
  2024.

\bibitem[de~Almeida and Thouless(1978)]{de1978stability}
Jairo~RL de~Almeida and David~J Thouless.
\newblock Stability of the sherrington-kirkpatrick solution of a spin glass
  model.
\newblock \emph{Journal of Physics A: Mathematical and General}, 11\penalty0
  (5):\penalty0 983, 1978.

\bibitem[Biscari and Parisi(1995)]{Biscari1995}
Paolo Biscari and G~Parisi.
\newblock Replica symmetry breaking in the random replicant model.
\newblock \emph{Journal of Physics A: Mathematical and General}, 28\penalty0
  (17):\penalty0 4697, 1995.

\bibitem[Diederich and Opper(1989)]{Diederich1989}
Sigurd Diederich and Manfred Opper.
\newblock Replicators with random interactions: A solvable model.
\newblock \emph{Physical Review A}, 39\penalty0 (8):\penalty0 4333, 1989.

\bibitem[M{\'e}zard et~al.(1987)M{\'e}zard, Parisi, and
  Virasoro]{mezard1987spin}
Marc M{\'e}zard, Giorgio Parisi, and Miguel~Angel Virasoro.
\newblock \emph{Spin glass theory and beyond: An Introduction to the Replica
  Method and Its Applications}, volume~9.
\newblock World Scientific Publishing Company, 1987.

\bibitem[Wu et~al.(2021)Wu, Mehta, and Schwab]{wu2021understanding}
Jim Wu, Pankaj Mehta, and David Schwab.
\newblock Understanding species abundance distributions in complex ecosystems
  of interacting species.
\newblock \emph{arXiv preprint arXiv:2103.02081}, 2021.

\bibitem[Ma(2020)]{ma2020testing}
Zhanshan~Sam Ma.
\newblock Testing the anna karenina principle in human microbiome-associated
  diseases.
\newblock \emph{Iscience}, 23\penalty0 (4), 2020.

\bibitem[Pigani et~al.(2024)Pigani, Mele, Campese, Ser-Giacomi, Ribera,
  Iudicone, and Suweis]{pigani2024}
Emanuele Pigani, Bruno~Hay Mele, Lucia Campese, Enrico Ser-Giacomi, Maurizio
  Ribera, Daniele Iudicone, and Samir Suweis.
\newblock Deviation from neutral species abundance distributions unveils
  geographical differences in the structure of diatom communities.
\newblock \emph{Science Advances}, 10\penalty0 (10):\penalty0 eadh0477, 2024.

\bibitem[Arnoulx~de Pirey and Bunin(2024)]{pirey-PRX}
Thibaut Arnoulx~de Pirey and Guy Bunin.
\newblock Many-species ecological fluctuations as a jump process from the brink
  of extinction.
\newblock \emph{Physical Review X}, 14\penalty0 (1):\penalty0 011037, 2024.

\bibitem[Tomasulo et~al.(2024)Tomasulo, Simionati, and Facchin]{tomasulo}
Antonietta Tomasulo, Barbara Simionati, and Sonia Facchin.
\newblock Microbiome one health model for a healthy ecosystem.
\newblock \emph{Science in One Health}, page 100065, 2024.

\bibitem[Franzosa et~al.(2019)Franzosa, Sirota-Madi, Avila-Pacheco, Fornelos,
  Haiser, Reinker, Vatanen, Hall, Mallick, McIver, et~al.]{franzosa2019gut}
Eric~A Franzosa, Alexandra Sirota-Madi, Julian Avila-Pacheco, Nadine Fornelos,
  Henry~J Haiser, Stefan Reinker, Tommi Vatanen, A~Brantley Hall, Himel
  Mallick, Lauren~J McIver, et~al.
\newblock Gut microbiome structure and metabolic activity in inflammatory bowel
  disease.
\newblock \emph{Nature microbiology}, 4\penalty0 (2):\penalty0 293--305, 2019.

\bibitem[Mars et~al.(2020)Mars, Yang, Ward, Houtti, Priya, Lekatz, Tang, Sun,
  Kalari, Korem, et~al.]{mars2020longitudinal}
Ruben~AT Mars, Yi~Yang, Tonya Ward, Mo~Houtti, Sambhawa Priya, Heather~R
  Lekatz, Xiaojia Tang, Zhifu Sun, Krishna~R Kalari, Tal Korem, et~al.
\newblock Longitudinal multi-omics reveals subset-specific mechanisms
  underlying irritable bowel syndrome.
\newblock \emph{Cell}, 182\penalty0 (6):\penalty0 1460--1473, 2020.

\bibitem[Zamponi(2010)]{zamponi2010}
Francesco Zamponi.
\newblock Mean field theory of spin glasses.
\newblock \emph{arXiv preprint arXiv:1008.4844}, 2010.

\bibitem[Virtanen et~al.(2020)Virtanen, Gommers, Oliphant, Haberland, Reddy,
  Cournapeau, Burovski, Peterson, Weckesser, Bright, et~al.]{virtanen2020scipy}
Pauli Virtanen, Ralf Gommers, Travis~E Oliphant, Matt Haberland, Tyler Reddy,
  David Cournapeau, Evgeni Burovski, Pearu Peterson, Warren Weckesser, Jonathan
  Bright, et~al.
\newblock Scipy 1.0: fundamental algorithms for scientific computing in python.
\newblock \emph{Nature methods}, 17\penalty0 (3):\penalty0 261--272, 2020.

\bibitem[Yoshino et~al.(2008)Yoshino, Galla, and Tokita]{yoshino2008rank}
Yoshimi Yoshino, Tobias Galla, and Kei Tokita.
\newblock Rank abundance relations in evolutionary dynamics of random
  replicators.
\newblock \emph{Physical Review E}, 78\penalty0 (3):\penalty0 031924, 2008.

\bibitem[Altieri and Franz(2019)]{altieri2019}
Ada Altieri and Silvio Franz.
\newblock Constraint satisfaction mechanisms for marginal stability and
  criticality in large ecosystems.
\newblock \emph{Physical Review E}, 99\penalty0 (1):\penalty0 010401, 2019.

\bibitem[Tikhonov and Monasson(2017)]{tikhonov2017}
Mikhail Tikhonov and Remi Monasson.
\newblock Collective phase in resource competition in a highly diverse
  ecosystem.
\newblock \emph{Physical review letters}, 118\penalty0 (4):\penalty0 048103,
  2017.

\bibitem[Guerra and Toninelli(2002)]{guerra2002}
Francesco Guerra and Fabio~Lucio Toninelli.
\newblock Quadratic replica coupling in the sherrington--kirkpatrick mean field
  spin glass model.
\newblock \emph{Journal of Mathematical Physics}, 43\penalty0 (7):\penalty0
  3704--3716, 2002.

\bibitem[Panchenko and Talagrand(2007)]{panchenko2007}
Dmitry Panchenko and Michel Talagrand.
\newblock On the overlap in the multiple spherical sk models.
\newblock 2007.

\bibitem[Dia et~al.(2016)Dia, Macris, Krzakala, Lesieur, Zdeborov{\'a},
  et~al.]{dia2016}
Mohamad Dia, Nicolas Macris, Florent Krzakala, Thibault Lesieur, Lenka
  Zdeborov{\'a}, et~al.
\newblock Mutual information for symmetric rank-one matrix estimation: A proof
  of the replica formula.
\newblock \emph{Advances in Neural Information Processing Systems}, 29, 2016.

\bibitem[De~Dominicis and Giardina(2006)]{dedominicis2006}
Cirano De~Dominicis and Irene Giardina.
\newblock \emph{Random fields and spin glasses: a field theory approach}.
\newblock Cambridge University Press, 2006.

\bibitem[Altieri et~al.(2016)Altieri, Parisi, and Rizzo]{altieri2016composite}
Ada Altieri, Giorgio Parisi, and Tommaso Rizzo.
\newblock Composite operators in cubic field theories and link-overlap
  fluctuations in spin-glass models.
\newblock \emph{Physical Review B}, 93\penalty0 (2):\penalty0 024422, 2016.

\end{thebibliography}

\end{document}